\documentclass[pre,superscriptaddress,showpacs,twocolumn,nobibnotes]{revtex4}
\usepackage{psfig}
\usepackage{amssymb}
\def\/{\over}
\def\<{\langle}
\def\>{\rangle}
\def\l<{\left\langle}
\def\r>{\right\rangle}
\def\({\left (}
\def\){\right )}
\def\[{\left [}
\def\]{\right ]}
\def\d{{\rm d}}
\def\i{{\rm i}}
\def\e{{\rm e}}
\def\mt{m_t}
\def\wt{{\mu_t}}
\def\nx{n_x}
\def\wx{{\nu_x}}
\def\Tr{{\rm tr}\,}
\def\x{\xi}
\def\t{\tau}
\def\E{{\cal E}}
\def\H{{\cal H}}

\def\T{{\cal T}}
\def\V{{\cal V}}
\def\mod{{\,\mbox{mod}\,}}
\def\figdir{EPS}
\def\ch{_{\rm ch}}
\begin{document}
\title{Directed Chaotic Transport in Hamiltonian Ratchets}
\author{Holger Schanz}
\email{holger@chaos.gwdg.de}
\affiliation{Max-Planck-Institut f\"ur Str\"omungsforschung und Institut
f{\"u}r Nichtlineare Dynamik der Universit{\"a}t G{\"o}ttingen,
Bunsenstra{\ss}e 10, D-37073 G\"ottingen, Germany}
\author{Thomas Dittrich}
\affiliation{Departamento de F\'isica, Universidad Nacional,
Santaf\'e de Bogot\'a, Colombia}
\author{Roland Ketzmerick}
\affiliation{Institut f\"ur Theoretische Physik, Technische
Universit{\"a}t Dresden, D-01062 Dresden, Germany} 
\date{\today}
\pacs{05.60.-k, 05.45.Mt}
\begin{abstract}
We present a comprehensive account of directed transport in
one-dimensional Hamiltonian systems with spatial and temporal
periodicity. They can be considered as Hamiltonian ratchets in the
sense that ensembles of particles can show directed ballistic
transport in the absence of an average force. We discuss
general conditions for such directed transport, like a mixed classical phase
space, and elucidate a sum rule that relates the contributions of
different phase-space components to transport with each other. We show
that regular ratchet transport can be directed against an external
potential gradient while chaotic ballistic transport is restricted to
unbiased systems. For quantized Hamiltonian ratchets we study
transport in terms of the evolution of wave packets and derive a
semiclassical expression for the distribution of level velocities
which encode the quantum transport in the Floquet band spectra. We
discuss the role of dynamical tunneling between transporting islands
and the chaotic sea and the breakdown of transport in quantum ratchets
with broken spatial periodicity.
\end{abstract}
\maketitle

\section{Introduction}

Hamiltonian systems with a mixed phase space remain a challenge within the
field of nonlinear dynamics, both classical and quantum. This is usually
attributed to the intricate, typically self-similar structure of phase space
in these systems. There exist, however, more tangible effects which also
require a coexistence of regular and chaotic dynamics but no particular fine
structure. A prominent example is directed transport: An elementary yet
decisive consequence of a mixed phase space is the existence of distinct
regions which support qualitatively different dynamics and do not communicate
with each other. Directed transport may arise locally in regular components of
phase space. As a consequence of a global sum rule, and in the absence of
certain symmetries, it can then be conferred to the chaotic component, as we
will show in this paper.

Chaotic transport in extended Hamiltonian systems is usually associated with
undirected diffusion: The width of the spatial distribution $\Delta x$ grows
with time as some power law $(\Delta x)^2\sim t^{\alpha}$ with $\alpha$
between $0$ and $2$. Only recently it has been discovered that even in the
absence of a mean external gradient, chaotic diffusion in driven Hamiltonian
systems can be accompanied by a directed drift. The corresponding ballistic
component of transport \cite{FYZ00} may surprise on first sight,
since a hallmark of chaos is the decay of all correlations including an
effective randomization of the velocity with time.  However, this implies only
that the mean velocity of a typical chaotic trajectory approaches an
asymptotic value which is characteristic of the chaotic phase-space region as
a whole. In the absence of additional symmetries there is no general reason
requiring this asymptotic mean velocity to be zero. 

In fact, as we shall argue in Sec.~\ref{sec:classical}, in systems with a
mixed phase space a sum rule requires chaotic transport to compensate for the
directed transport possibly occurring in regular phase-space regions
\cite{D+00,S+01}. An important conclusion (Sec.~\ref{sec:barriers}) is that
the ballistic chaotic transport has nothing to do with internal structures of
a chaotic phase-space component such as cantori or other partial transport
barriers. All these complicated substructures, leading, e.g., to L\'evy walks
and anomalous diffusion in Hamiltonian ratchets \cite{FYZ00,DF01,D+02}, need
not be considered in detail in order to understand that ballistic transport
dominates for long times.

Deterministic ballistic transport due to a dynamical restriction of
trajectories to certain phase-space regions has been observed before in
dissipative systems \cite{JKH96}, where phase-space volume is contracting
with time. This mechanism is close to the concept of stochastic ratchets
(Brownian motors), i.e., systems that generate directed motion from
non-equilibrium noise \cite{Feynman,JAP97,Rei02}. The analogy suggests to
speak of deterministic ratchets.

Throughout this paper, we disregard dissipation. Its absence, however, not
only renders it more difficult to achieve directed transport, since the
natural time arrow determined by dissipation is lost and has to be replaced by
other mechanisms breaking time-reversal invariance. It even becomes a subtle
task to define a ratchet in a Hamiltonian framework in the first
place. Trajectories can maintain a memory of their
initial velocity for an infinite time. Therefore a precise definition of a {\em
Hamiltonian ratchet} is not completely straightforward: The mere fact that
in unbiased systems directed transport can exist and survive for infinite
time is trivial, just take a free particle with some non-zero initial
velocity $v_0\ne 0$. In this sense every extended Hamiltonian system
would be a ratchet. 

Due to velocity dispersion an ensemble of free particles will also spread
ballistically, i.e., as fast as its center of mass is transported. On the
other hand, as pointed out above, there exist Hamiltonian systems where
transport is ballistic, but the spreading is not. They are characterized by a
locking of the average velocity to a (non-zero) value which does not depend on
the precise initial conditions as long as they are restricted to some finite
phase-space region. For the purpose of the present paper we regard this
property as the definition of a Hamiltonian ratchet.

\begin{figure}[!b]
 \centerline{
\begin{tabular}{ll}
\sf (a)&\sf (b)\\[-3mm]
\hspace*{3mm}  
\psfig{figure=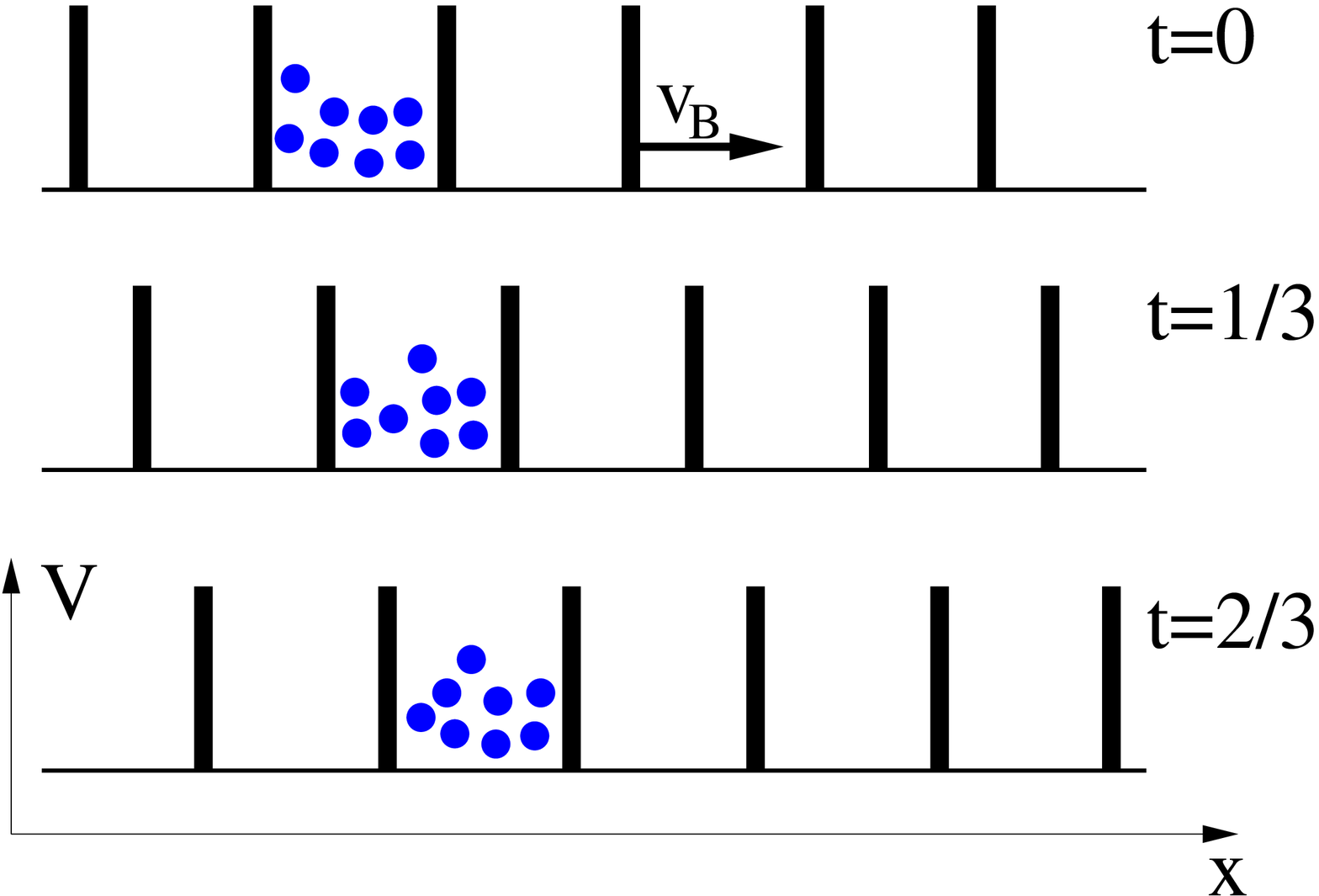,height=30mm}
&
\psfig{figure=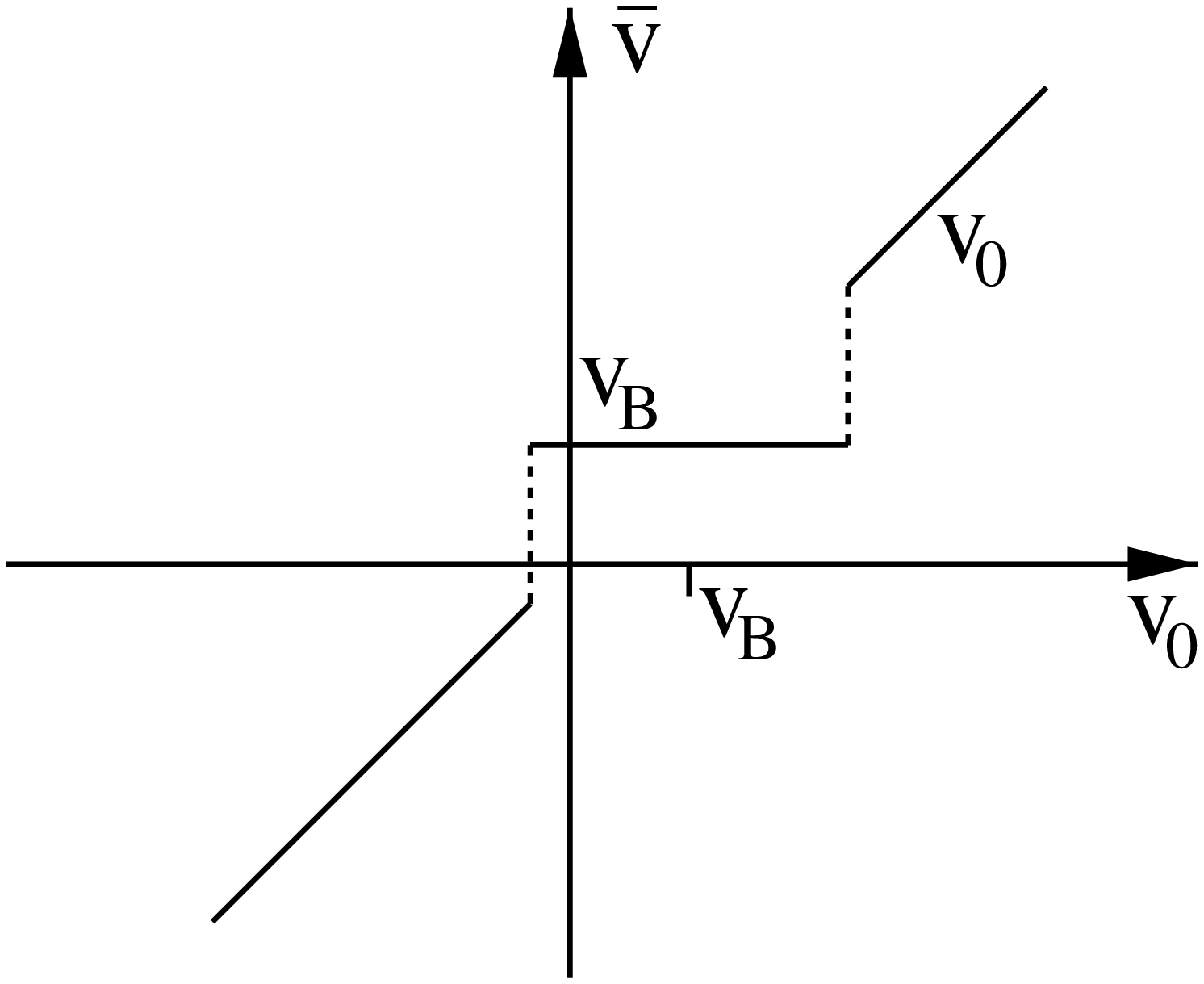,height=30mm}
\end{tabular}
}
 \caption{\label{moving_comb}
   (a) A trivial example for a Hamiltonian ratchet is a periodic potential
   which is moving at a constant velocity $v_{\rm B}>0$ such that
   $V(x,t)=\tilde V(x-v_{\rm B}t)$.  The system is integrable, since it is
   time independent in the comoving reference frame. Despite the conveyor-belt
   construction it is also unbiased, since the average force in a periodic
   potential is always zero.  (b) shows the dependence of the asymptotic mean
   velocity $\overline v$ on the initial velocity $v_{0}$ under the assumption
   that the potential is non-zero only in negligibly small intervals.
   Particles with initial velocity close to $v_{\rm B}$, namely for
   $m(v_0-v_{\rm B})^2/2<V_{\rm max}$, are trapped inside one well of the
   potential and have an asymptotic velocity $\overline v=v_{\rm B}$
   independent of the precise initial conditions.}
\end{figure}

Even with this restriction it is possible to construct cases one would
qualify as trivial realizations of directed transport: In the
integrable system sketched in Fig.~\ref{moving_comb}, for example,
transport appears to be achieved by a mere change of frame. For the
sake of simplicity of the definition we do not attempt to formally
exclude such cases. In what follows, however, we concentrate on
extended systems with a mixed phase space where one has to understand
the interplay between regular and chaotic transport. 

Unless symmetries of the driving potential prevent it \cite{FYZ00},
Hamiltonian ratchets as defined above lead without average force to a non-zero
mean velocity of an ensemble of particles which were initially at rest. The
same applies also to the ratchets described in \protect\cite{M+02}, although
there is no velocity locking and ensembles of particles do spread
ballistically. These systems are based on a mechanism that is different from
the models discussed in \protect\cite{FYZ00,DF01,D+02,GH00,D+00,S+01} and the
present paper, and we will not consider them here.

For Hamiltonian ratchets under the influence of an average force we show in
Sec.~\ref{sec:bias} that uphill regular transport is possible. In contrast,
even an infinitesimal average force destroys the chaotic drift and leads to
downhill acceleration.

It comes as a rather unexpected finding that Hamiltonian ratchets have
applications on macroscopic, even geophysical scales where apparently
friction prevails \protect\cite{Neg98}. Indeed, in hydrodynamics, even in
the presence of dissipation, restricting the description to position space
results in a Hamiltonian form of the evolution equations if only the fluid is
\emph{incompressible}. Specifically, in geophysical applications, a
periodic potential reflects the periodic boundary conditions on Earth with
respect to longitude, while an asymmetry in the transverse coordinate is
implied by the dependence of the Coriolis force on latitude.

Going in the opposite direction, Hamiltonian ratchets are to find
applications on scales where quantum effects become important. For
example, in semiconductor nanostructures employed to investigate
solid-state ratchets \cite{L+99,H+01} such effects were observed.  A
Hamiltonian ratchet with negligible dissipation can be realized on this basis
if the structure size is further decreased, such that electronic motion occurs
in the ballistic regime. But this will even enhance quantum
corrections. 

In Ref.~\cite{S+01} it was concluded that quantum Hamiltonian ratchets can
work if classical and quantum system are both spatially periodic such that the
quantum system has a band spectrum. Detailing our findings, we will show in
Section \ref{sec:shorttime} that quantum transport relies on the semiclassical
correspondence between the dynamics of wave packets and that of classical
distributions in phase space: As long as a wave packet, started in the chaotic
region of phase space, say, remains predominantly restricted to this region,
it will be transported with the classical mean chaotic velocity. Such
quantum-classical correspondence can be attributed to the existence of
different types of bands in the spectrum, with eigenfunctions concentrating
semiclassically on different invariant sets of classical phase space. Since
this mechanism crucially depends on classical phase-space structures, it
cannot be captured using a single- (or few-) band picture. Therefore our
results are not at variance with the absence of transport demonstrated within
such an approximation \cite{GH00}.

However, also in the semiclassical regime non-classical processes
like tunneling are possible which allow transitions between invariant
sets of classical phase space. In Section \ref{sec:longtime} we will address
the question why this is compatible with quantum transport unlimited in
time. Only when the exact periodicity of the quantum system is destroyed,
the eigenfunctions governing the long-time dynamics ignore classical
phase-space structures \cite{H+02} such that ratchet transport becomes a
transient phenomenon. We shall deal with this case in Section
\ref{sec:disorder}.

In our conclusions (Sec.~\ref{sec:disc}) we discuss in particular various ways
of breaking the translation invariance of Hamiltonian ratchets and how this
affects transport.

\section{Classical Hamiltonian ratchets}
\label{sec:classical}

\subsection{The Hamiltonian of the extended system}
\label{Hamil}

We consider Hamiltonian systems in one dimension which are {\em periodic} and
{\em unbiased} in the sense specified below. The Hamiltonian is of the form
\begin{equation}\label{Hamiltonian}
H(p,x,t)=T(p)+V(x,t)\,,
\end{equation}
where $x$ and $p$ are the canonically conjugate position and momentum
and $T(p)$ and $V(x,t)$ denote kinetic and potential energy,
respectively.  

We require that the {\em dynamics} be invariant under integer translations of space
or time and use dimensionless variables in which both periods are unity, i.e.,
we assume the following property: For any trajectory $x(t)$ with initial
conditions $x(t_{0})=x_{0}$, $p(t_{0})=p_{0}$ and any other trajectory $\tilde
x(t)$ with $\tilde x(t_{0}+n)=x_{0}+m$, $\tilde p(t_{0}+n)=p_{0}$ we have
$\tilde x(t+n)=x(t)+m$ for all $t$.

In the simplest case this is realized by $T(p)=p^2/2$ and a spatially and
temporally periodic potential
\begin{equation}\label{PeriodicPotential}
V(x,t+1)=V(x+1,t)=V(x,t)\,,
\end{equation}
but this is not a necessary condition: If the potential contains an additional
term $f(t)\,x$ we have $V'(x+1,t)=V'(x,t)$ only, where $V'=\d V/\d x$.
Nevertheless, discrete translation invariance may be satisfied for the
dynamics, see Section~\ref{sec:bias} for an example.

We shall refer to the system as unbiased, if the force $-V'$ averaged over
space and time vanishes
\begin{equation}\label{unbiased}
\int_{0}^{1}\d x\,\int_{0}^{1}\d t\, V'(x,t)=0\,.
\end{equation}
In Section~\ref{sec:minmod} we will also consider systems where the kinetic
energy is a periodic function of $p$ such as $T(p)=\cos 2\pi p$ for electrons
in a Bloch band. As we shall see, such systems are always unbiased. 


\subsection{The phase space of a unit cell}\label{sec:psec}

Instead of the extended system represented by (\ref{Hamiltonian}), the
discrete translation invariance allows to consider an auxiliary system
restricted to a single {\em unit cell} by imposing periodic boundary
conditions at $x=1$, $t=1$. Since in this paper both representations appear
in parallel, we use different symbols $\x\equiv x\,\mod 1$ and $\t\equiv
t\,\mod 1$ for the cyclic variables of the unit cell.

It is a standard technique for driven systems \cite{CM89} to treat time like a
spatial coordinate such that a one-dimensional time-dependent system is
mapped to a formally time-independent problem in two dimensions. For the unit
cell the Hamiltonian obtained in this way is
\begin{equation}\label{ucham}
\H(\x,p,\t,\E)=T(p)+V(\xi,\t)+\E\,,
\end{equation}
where $\E$ is canonically conjugate to $\t$. This ensures
$\dot\t=\partial\H/\partial \E=1$. Since $\H$ is a conserved quantity,
$-\E(t)$ can be interpreted as the energy $\Delta H$ which the system has
gained from the driving up to time $t$. Moreover it becomes
clear that the dynamics is restricted to a three-dimensional ``energy shell",
$\H=$ const., which is spanned by the variables $\x$, $p$ and $\t$ ($\E$ is a
function of these three variables and the constant $\cal H$).

\begin{figure}[!b]
 \centerline{\psfig{figure=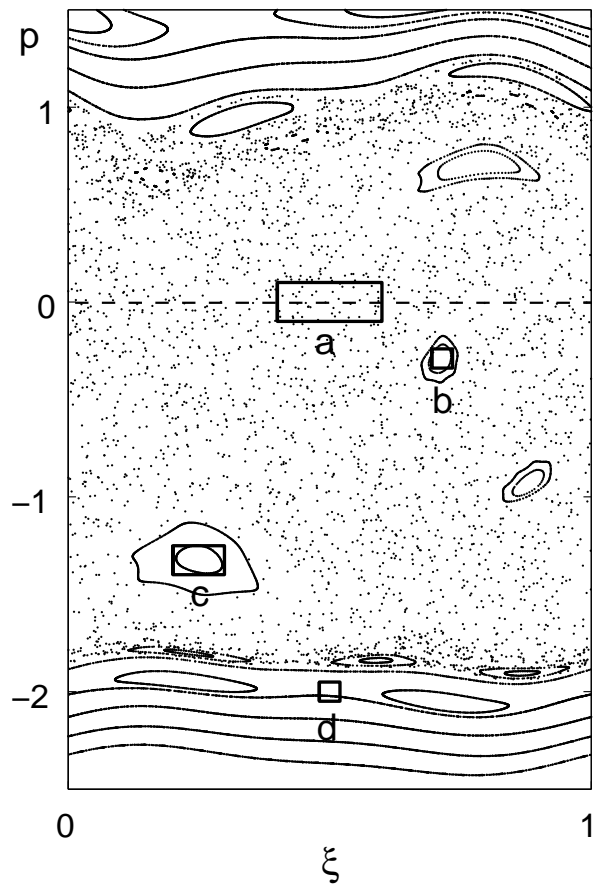,width=8cm}}
 \caption{\label{psec_fyz} Typical stroboscopic Poincar\'e section $\t=0$ for a
   Hamiltonian ratchet with non-contractible KAM tori, main chaotic sea and
   regular islands.
   The lettered rectangular regions support initial distributions of particles
   for which the corresponding velocity
   distributions are shown in Fig.~\protect\ref{distributions}.}
\end{figure}

The dimensionality can be reduced further by considering Poincar\'e surfaces of
section at some constant $\t$ which eliminates the trivial flow in
$\t$-direction.
In the following, we shall discuss the main features of such stroboscopic
surfaces of section, relevant for transport in Hamiltonian ratchets. For the
moment we restrict the discussion to smooth potentials in the sense of the
Kolmogorov-Arnol'd-Moser (KAM) theorem \cite{Ott93} and take as an example the
Hamiltonian
\begin{equation}\label{fyzham}
H(p,x,t)={p^2\over 2}+V_{0}(x)+x\,V_{1}(t)
\end{equation}
with
\begin{equation}
V_{0}(x)={1\over 5.76}[\sin(2\pi x)+0.3\sin(4\pi x+0.4)]
\end{equation}
and
\begin{equation}
V_{1}(x)=-{\pi\over 5.76}[4.6\,\sin(2\pi t)+2.76\,\sin(4\pi t+0.7)]\,.
\end{equation}
This corresponds to the parameter set (3) of Fig.~1 in
Ref.~\protect\cite{FYZ00} when the spatial and the temporal period are scaled
to unity.

The stroboscopic Poincar\'e section for this model is shown in
Fig.~\ref{psec_fyz}.  We can distinguish the following three types of motion,
each corresponding to a characteristic signature in phase space and transport:

(i) At high kinetic energies the ratchet potential can be considered a small
perturbation acting on a free particle. For this integrable limit the
trajectories are confined to invariant surfaces in phase space which have the
topology of a torus. These tori are labeled by the conserved value of the
momentum $p$ and parameterized by the cyclic variables $\x$ and $\t$.  In the
$(\x,p)$ plane of the stroboscopic Poincar\'e section the tori would
consequently appear as horizontal lines.

The KAM theorem predicts the fate of a torus under a small perturbation. It
depends on whether its {\em winding number} $w$ is rational or not. The
winding number is the ratio between the angular velocities along the two
independent cyclic coordinates spanning the torus. In the present case, one of
these coordinates is the time $\t$ and the corresponding angular velocity
is unity by definition. For the other coordinate $\x$, the angular
velocity on the torus is equal to the transport velocity in the extended
system, measured in spatial unit cells per time period, so that $w=\overline
v$. 

Almost all tori have irrational winding numbers and, according to the KAM
theorem, most of them survive an infinitesimal perturbation. This is visible
in Fig.~\ref{psec_fyz} at high $|p|$ where we observe lines in the
stroboscopic Poincar\'e section which extend across the unit cell. Although
the lines are deformed by the potential they represent intact tori of
regular motion with irrational winding number (transport velocity). Motion
proceeds on these tori in the initial direction, without turning points.
As these tori cannot be continuously contracted to a point we will call them
{\em non-contractible}.

(ii) Tori with rational winding number 
\begin{equation}\label{wz}
w=\wx/\wt\,, 
\end{equation}
$\wx$, $\wt$ integer, which comprise a set of measure zero, are
destroyed under an
infinitesimal perturbation. Details of this effect are described by
the Poincar\'e-Birkhoff-Theorem \cite{Ott93}. Together with a small
neighborhood, a rational torus decays to a chaotic layer embedding new
tori of regular motion. These tori have a different topology, however:
They are contractible and appear as a set of $\wt$ regular islands in the 
stroboscopic Poincar\'e section. The respective centers of the
islands are formed by a single elliptic periodic orbit with period
$\wt$, i.e., this orbit has $\wt$ distinct intersections with the stroboscopic
Poincar\'e section before it starts repeating, shifted by $\wx$ unit cells
in the extended space. In Fig.~\ref{psec_fyz}, we therefore observe chains
of regular islands which are sequentially traversed by a trajectory. The
average velocity of the central periodic orbit and of all trajectories
inside the island corresponds to the rational winding number $\overline
v=w$ of the destroyed torus. If $\overline v \ne 0$ we speak of a
\emph{transporting island}. 

(iii) The chaotic regions surrounding the island chains at high $|p|$ are
too small to be visible in Fig.~\ref{psec_fyz}. With
increasing perturbation, however, the chaotic regions grow and may 
coalesce. In the vicinity of $p=0$ the effective perturbation is
strongest. As a result a large {\em chaotic sea} develops. With
increasing resolution we find more and more islands embedded in this sea
and more and more chains of transporting islands interrupting the strips
where the intact KAM tori reside.  Such islands need not be remnants of
rational tori in the undriven system---they can appear and disappear at
some finite value of the driving potential as a result of bifurcations of
periodic orbits. Still, their transport velocity
must also be given by a rational winding number. 

Conversely, we find more and more small chaotic regions located
within the regular islands. Since they are confined to the
phase-space region demarked by the outermost intact torus encircling the island,
they share the same average velocity $\overline v = w$, where $w$ is the
winding number of the island.

The phase-space regions enumerated above are most adequately discussed in
terms of \emph{invariant sets}: a subset of phase space invariant \emph{as a
  whole} under the dynamics, irrespective of any reshuffling possibly
occurring inside. For example, any regular torus in the three-dimensional
phase space of the unit cell is invariant under the dynamics. Trajectories
initialized on the torus do not leave it and vice versa. This invariant
two-dimensional surface separates the remaining phase space into two invariant
sets of non-zero measure. Moreover, any region in phase space confined by a
number of tori is an invariant set of the dynamics. In particular, this
applies to the chaotic sea, which is bounded from below and above by two
non-contractible KAM tori and by the outermost tori of the embedded regular
islands.

For our purpose the limitation of chaotic trajectories to a compact region of
phase space will be crucial. In the example discussed above this is a
consequence of the KAM scenario. In systems where the KAM theorem is not
valid our theory applies as long as there is another mechanism leading to
a compact chaotic phase-space component. An example of this type will be
discussed in Section~\ref{sec:minmod}.


\subsection{Velocity distribution}

Although the system (\ref{ucham}) is restricted to a single unit cell, it
contains the complete information about transport in the extended system
(\ref{Hamiltonian}). The velocity $v=\d H/\d p=T'(p)$ along a trajectory is
the same in both cases provided the initial conditions are equivalent, i.e.,
$\x_0=x_0\,\mbox{mod}\,1$ at $t=0$. Therefore the velocity is the appropriate
quantity to connect transport in the extended system to the unit cell and we
describe transport in terms of the velocity distribution for an ensemble
of particles. An ensemble is specified by a normalized initial distribution
$\rho_{0}(\x_0,p_0,\t_0)$ in the phase-space unit cell. The variable $\t_0$ is
part of the initial conditions since it matters at which phase of the driving
force a trajectory was started. It can indeed be physically meaningful to
consider ensembles for which $\t_0$ is not sharp, for example to model a
situation where particles continuously enter the system.

\begin{figure}[!b]
 \centerline{
  \psfig{figure=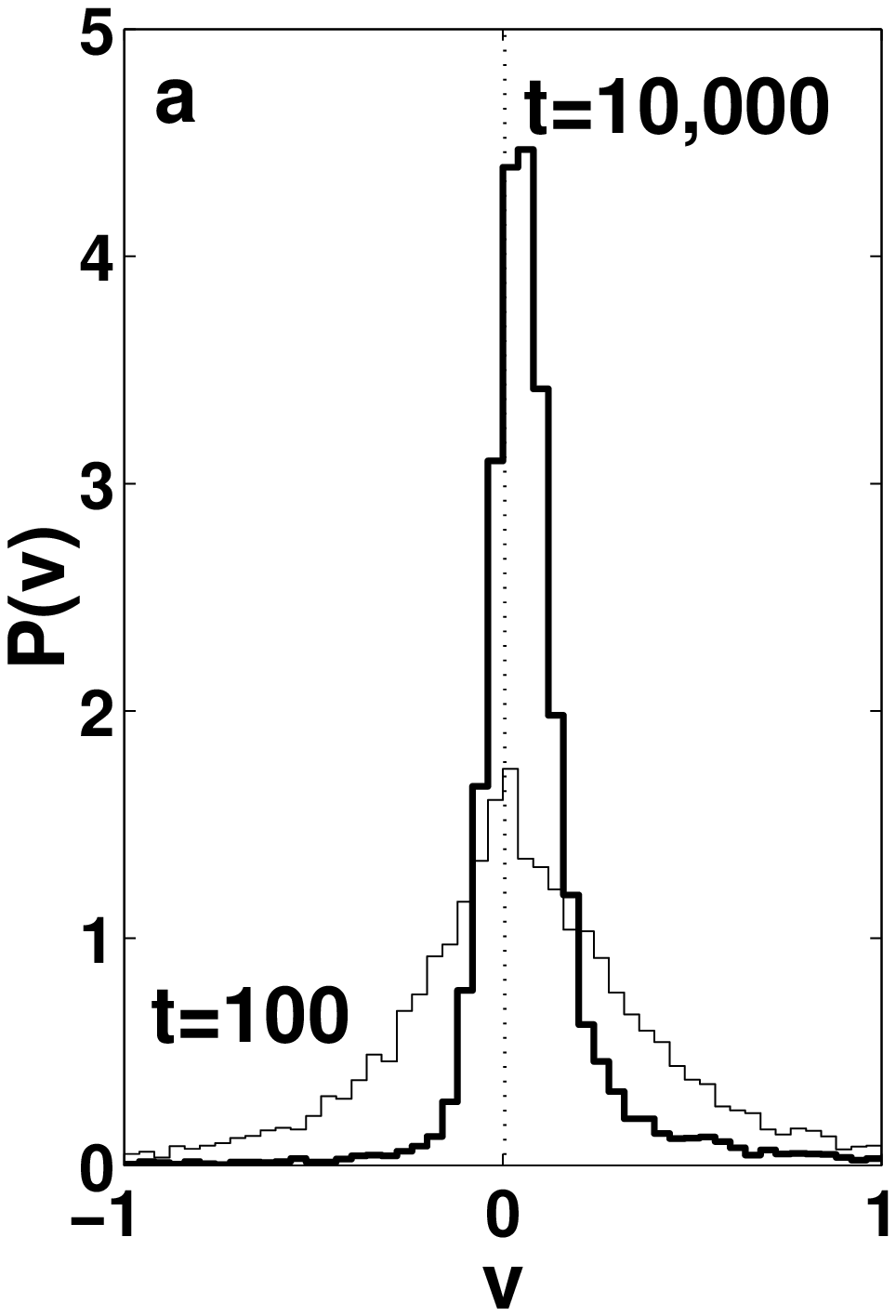,width=42mm}
  \psfig{figure=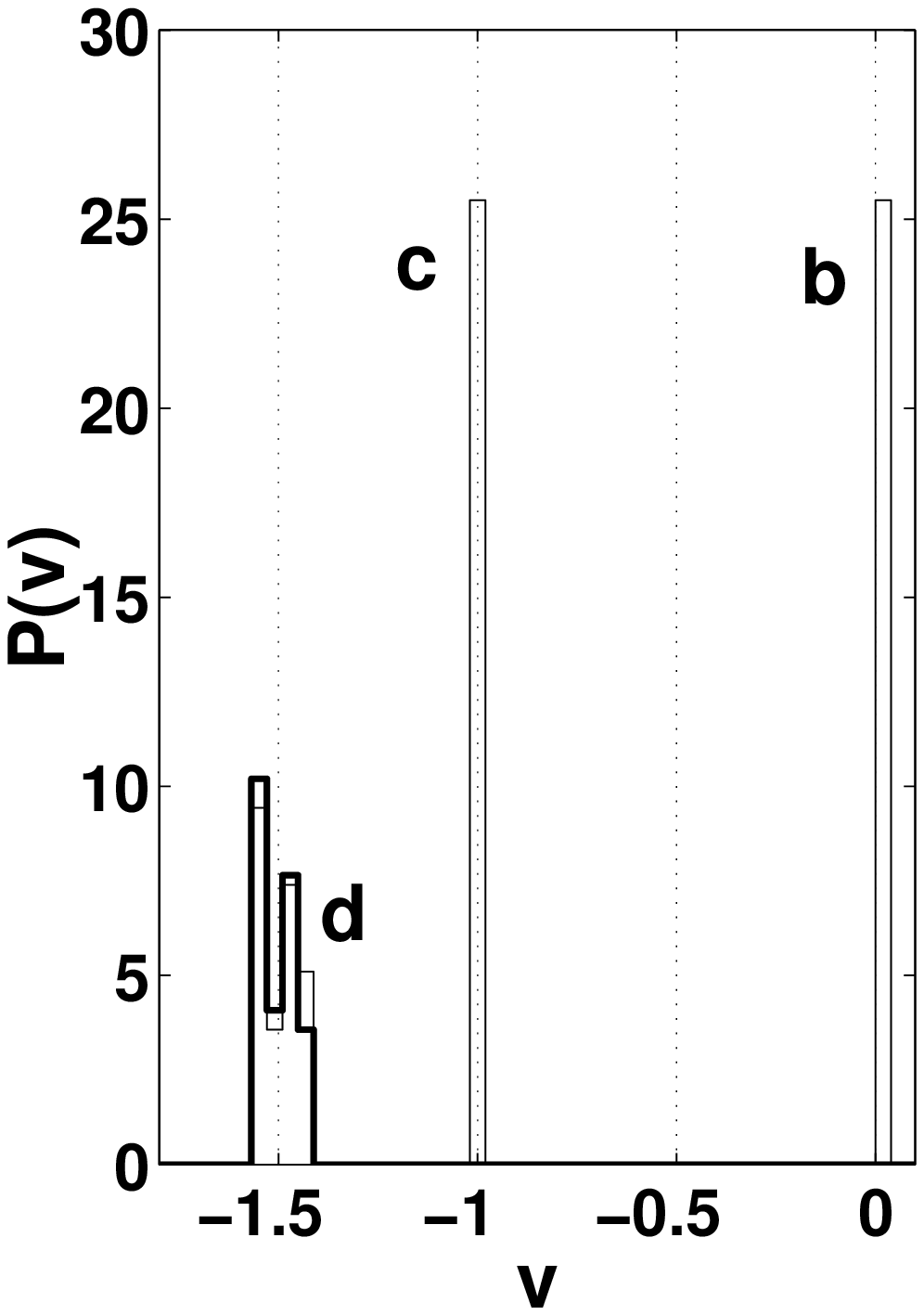,width=43mm}
 }
 \caption{\label{distributions}
   Distribution of time-averaged velocities, Eq.~(\protect\ref{velodist}), for
   four different initial distributions a--d (see Fig.~\protect\ref{psec_fyz}).
   The chaotic distribution (a) was sampled by 10,000 trajectories, while for
   each of the regular distributions (b--d) only 100 trajectories were used.
   For each trajectory the velocity was averaged up to $t=100$ and the
   resulting distributions are displayed with solid lines (a--d). For (a), (d)
   also the distributions at $t=10,000$ are shown (bold lines).}
\end{figure}

For any ensemble $\rho_0$ and time $t$ we define the time-averaged velocity
distribution as
\begin{eqnarray}\label{velodist}
P_{\rho_0,t}(v)&=&{1\/t}\int_{0}^{t}\d t'\,\int_{-\infty}^{+\infty}\d
p_0\,\int_{0}^{1}\d \xi_{0}\,\int_{0}^{1}\d\t_0\, \nonumber\\
&&
\times
\rho_{0}(\xi_0,p_0,\t_0)\,\delta(v-T'(p_{t';\xi_0,p_0,\t_0}))
\end{eqnarray}
with the normalization $\int\d v\,P(v)=1$. 
If we consider an ensemble in the extended
system, initially localized at $x=0$, then at a later time $t$ its spatial
distribution will be given in terms of the velocity distribution by
$\rho_{t}(x)=t^{-1}\,P_{\rho_0,t}(x/t)$. For long times the center of
mass moves with the mean velocity 
\begin{equation}\label{velomean}
v_{\rho_0}=\int_{-\infty}^{+\infty}\d
v\,v\,P_{\rho_0,\infty}(v)\,,
\end{equation}
where the existence of $P_{\rho_0,\infty}(v)=\lim_{t\to\infty}P_{\rho_0,t}(v)$
is assumed.  

The behaviour of the velocity distribution is qualitatively different for
initial distributions $\rho_{0}$ which are restricted to different invariant
sets of the phase space. This is demonstrated in Fig.~\ref{distributions}.
We used as initial distributions the characteristic functions
$\chi_{a,b,c,d}$ of the rectangles marked in Fig.~\ref{psec_fyz},
approximated by a large number of trajectories with initial conditions
distributed randomly inside the corresponding region.

In the simplest case, $\rho_{0}$ has support inside a regular island
(distributions b, c in Fig.~\ref{distributions}).  According to the
last section, the average velocity of all trajectories inside an
island is equal to the winding number $w$ of the island.  Consequently we have
\begin{equation}\label{nospread}
P_{\rho_0,\infty}(v)=\delta(v-w)
\end{equation}
and observe sharp peaks in Fig.~\ref{distributions}b, c whose width is within
the bin size of the histogram already at $t=100$.  Fig.~\ref{distributions}c
is an example for a transporting island, $w\ne 0$.  Any distribution
$\rho_{0}$ initialized inside this island will be transported ballistically
with velocity $w=-1$. At the same time the width of the distribution does not
grow ballistically. As stated in the introduction, we consider this behaviour
as the defining property of a Hamiltonian ratchet.

For an ensemble initialized in the chaotic sea (Fig.~\ref{distributions}a) the
situation is similar. Although here the velocity distribution shows an
appreciable width at finite times, the comparison of $t=100$ and $t=10,000$
suggests that this width goes to zero as $t\to\infty$. We can explain this
behaviour using the concept of ergodicity. Ergodicity means that for any
function defined on phase space and for almost all trajectories the time
average along the trajectory coincides with an average over the accessible
phase space. It is usually assumed that this property applies to the chaotic
components of systems with a mixed phase space, although proofs of such a
statement can be given only in exceptional situations \cite{Bun01}.  For our
purpose we can use the velocity $v=T'(p)$ as the function on phase space and
obtain for any non-singular initial distribution inside the chaotic
sea, such as the rectangular region of Fig.~\ref{psec_fyz}a,
\begin{equation}\label{chvelodist}
P_{\rho_0,\infty}(v)=\delta(v-v_{\rm ch})
\end{equation}
with the mean chaotic velocity
\begin{equation}\label{vch}
v_{\rm ch}=\V_{\rm ch}^{-1}
\int_{\rm ch}\d \t\,\d \x\,\d p\; T'(p)\,.
\end{equation}
The phase-space integral extends here over the whole chaotic sea of the
spatio-temporal unit cell, and $\V_{\rm ch}=\int_{\rm ch}\d \t\,\d \x\,\d p$
denotes its volume.  

In the following section we shall discuss a method to evaluate Eq.~(\ref{vch}).
For the moment it suffices to say that, in the absence of specific symmetries,
there is no general reason to expect
that the chaotic velocity predicted by this equation is zero. Therefore,
also the chaotic sea provides an example for Hamiltonian ratchet transport.

For both, regular islands and chaotic components the asymptotic velocity
distribution is a $\delta$-function which does not depend on the precise
location of the initial phase-space distribution within the invariant set. The
velocity distribution obtained from a region with surviving non-contractible
KAM tori shows a fundamentally different behaviour, analogous to the case of a
free particle: it maintains a finite width for $t\to\infty$ and a complicated
internal structure (distribution d in Fig.~\ref{distributions}). Moreover, the
detailed properties of the asymptotic velocity distribution depend on the
precise shape and location of the initial ensemble. Hence, according to our
definition, non-contractible tori do not show ratchet-like transport.


\subsection{Transport for invariant sets and sum rule}
\label{sec:clsumrule}

There is an interesting reformulation of Eq.~(\ref{vch}) which allows to
calculate the chaotic mean velocity in terms of regular trajectories only
\cite{S+01}. For any subset $M$ of the unit cell, we define its
contribution to {\em transport}, $\T_{M}$, as phase-space volume times
average velocity,
\begin{eqnarray}\label{transport}
\T_{M}&=&\V_{M}v_{M}
\nonumber\\&=&
\int\d\x\,\d p\,\d\t\;\chi_{M}(\x,p,\t)T'(p)\,,
\end{eqnarray}
where $\chi_{M}(\x,p,\t)$ is the characteristic function of $M$.  Note that in
this definition $M$ is not necessarily an invariant set.  However, if $M$
denotes either the chaotic sea or a regular island, the phase-space averaged
velocity $v_{M}$ can be identified with the asymptotic mean velocity of almost
all trajectories inside the invariant set, as described in the
previous section. 

Transport has to be distinguished from the familiar concept of \emph{current}
which refers to the probability flow that passes per unit time through a
surface dividing phase space. Here we are interested in transport along the
$x$-direction. Therefore we consider the current at a point $\xi_{0}$.  The
value of the current depends on the position $\xi_{0}$ and the time $\tau_{0}$
where it is measured. In terms of the density $\rho_{\tau}(\xi,p)$, it
is given as
\begin{equation}
I(\xi_{0},\tau_{0})=\int_{-\infty}^{+\infty}\d p\,\rho_{\tau_{0}}(\xi_{0},p)\,T'(p)\,.
\end{equation}
In order to relate this current to the transport of an invariant set $M$,
Eq.~(\ref{transport}), we substitute the density of the invariant measure
\begin{equation}\label{dense}
\rho_{\tau}(\xi,p)={\chi_{M}(\xi,p,\tau)\/A_{M,\tau}}
\end{equation}
where $A_{M}$ denotes the area of $M$ in a stroboscopic Poincar\'e
section. Integration of the density over one period of the driving leads to
the time-averaged current of $M$ at $\xi_{0}$,
\begin{equation}
\overline I_{M}(\xi_{0})={1\over
  A_{M}}\int_{0}^{1}\d\tau\int_{-\infty}^{+\infty}\d p\,\chi_{M}(\xi_{0},p,\tau)\,T'(p)
\end{equation}
where we have used the conservation of phase-space area in Hamiltonian
systems, $A_{M,\tau}=A_{M}$. An additional integration over
$\xi_{0}$ yields the relation between current in $x$-direction and transport
\begin{equation}
\T_{M}=A_{M}\overline I_{M}\,.
\end{equation}
Here we have used that the time-averaged current is independent of the position
$\xi_{0}$, as implied by the continuity equation for the invariant
measure. Note that for this reason we could in principle define
transport also without the $\xi$-integration.

By choosing the density as in Eq.~(\ref{dense}) and weighting the contribution
of each invariant set $M$ by its area $A_{M}$, we achieve that the resulting
quantity, transport, is additive. Namely,  with the
definition (\ref{transport}), we have for the
union of two or more disjoint sets, i.e., for $M=\bigcup_{i} M_{i}$, with
$M_i \cap M_j =
\emptyset$ for all $i \neq j$,
\begin{equation}\label{SumRule}
\T_{M}=\sum_{i}\T_{M_{i}}\,.
\end{equation}
We will apply this {\em sum rule} for transport to the layer in phase
space which contains the chaotic sea and the embedded regular islands. It
is bounded from below and above by two KAM tori. For simplicity we assume
that they can be represented by two functions $p_{\rm u/l}(\x,\t)$. We
find from Eq.~(\ref{transport})
\begin{eqnarray}\label{tlayer}
\T_{\rm layer}&=&\int_{0}^{1}\d\x\,\int_{0}^{1}\d\t\,
\int_{p_{\rm l}(\x,\t)}^{p_{\rm u}(\x,\t)}\d p\;T'(p)
\nonumber\\&=&
\int_{0}^{1}\d\x\,\int_{0}^{1}\d\t\,[T(p_{\rm u}(\x,\t))-T(p_{\rm l}(\x,\t))]
\nonumber\\&=&
\<T\>_{\rm u}-\<T\>_{\rm l}\,,
\end{eqnarray}
i.e., the transport of the layer is simply given by the kinetic energy
$T$, averaged over the two bounding KAM tori. In short, since the
underlying phase-space distribution $\chi_{M}(\xi,p,\tau)$ is flat,
the transport is determined by the outline defining the subset $M$. This
applies to \emph{any} subset of phase space confined by two non
contractible tori. 

On the other hand, according to Eq.~(\ref{SumRule}) the transport of
the stochastic layer is equal to the contributions from the invariant
manifolds it comprises
\begin{equation}\label{SumRule2}
\<T\>_{\rm u}-\<T\>_{\rm l}=\V_{\rm ch} v_{\rm ch}+
\sum_{i}\V_{i} v_{i}\,.
\end{equation}
Eq.~(\ref{SumRule2}) can be used to predict the chaotic transport velocity. In
practice this works as follows:

(i) In the stroboscopic Poincar\'e section we determine the location of
the limiting KAM tori $p_{\rm u/l}$ and the location of the limiting tori
of all major regular islands $i$ together with their winding numbers
$w_{i}$.

(ii) In order to determine the phase-space volumes entering
Eq.~(\ref{SumRule2}) it is in fact sufficient to know the areas in the
stroboscopic Poincar\'e section: The Liouville theorem applied to the
time-dependent Hamiltonian Eq.~(\ref{Hamiltonian}) \protect\cite{Ott93}
ensures that such an area is conserved by the dynamics. The
three-dimensional volume within the phase space of the unit cell is simply the
area at any given moment in time, multiplied by the temporal period
$\V=A\times 1$. Areas in the Poincar\'e section are determined by
approximating the corresponding invariant manifold by a polygon with corners
obtained from running a trajectory on the outermost torus.
Numerically, an approximation to this torus can be found by zooming
into the Poincar\'e section. 

(iii) The kinetic-energy averages $\<T\>_{\rm u,l}$ over the bounding
KAM tori are 
obtained by sampling a torus with a long trajectory, and determining the
integrals Eq.~(\ref{tlayer}) numerically. Note that this is not equivalent to
a time average over such a trajectory as the invariant density on the torus is
not constant.

(iv) Putting all the information together we find
\begin{equation}\label{vch-from-sr}
v_{\rm ch}={\<T\>_{\rm u}-\<T\>_{\rm l}-
\sum_{i}A_{i}w_{i}\over A_{\rm layer}-\sum_{i}A_{i}}
\end{equation}
Compared to the above procedure, the straightforward method of
determining the chaotic transport velocity by running a very long
trajectory has the disadvantage that its accuracy is hard to control. The
trajectory must be long enough to sample the chaotic phase-space component
ergodically, and there is no way to tell from a single trajectory whether
this has been achieved with sufficient accuracy. The reason is that the
chaotic  component typically contains partial barriers (cantori), which
may appear closed in a simulation over finite time. The error made by
ignoring the phase-space region behind the partial barrier can in
principle be arbitrarily large. Also the converse error is possible:
For long simulations the accumulating numerical inaccuracy may drive a
chaotic trajectory beyond an intact KAM torus. By using a stroboscopic
Poincar\'e section such errors are substantially reduced. In the picture
obtained from many relatively short trajectories, sampling the entire
phase space, one can judge if there are two nearby chaotic regions which may
actually form a single invariant set. It is then sufficient to increase
the resolution selectively in a small portion of phase space, which is
possible with relatively small computational effort.


\subsection{Chaotic transport and L\'evy walks}
\label{sec:barriers}

Equation (\ref{vch-from-sr}) shows that the basic mechanism underlying
chaotic ratchet transport is the existence of KAM tori and regular islands
which prevent a chaotic trajectory from sampling the whole classical phase
space. Unless there are special symmetries, the velocity average over the
chaotic sea is generically non-zero and it is determined solely by the
boundaries of this invariant set. Besides ergodicity, no reference to any
details of the dynamics within the chaotic set is needed to explain and
quantitatively predict the observed asymptotic chaotic transport velocity.

\begin{figure}[!b]
\centerline{
  \psfig{figure=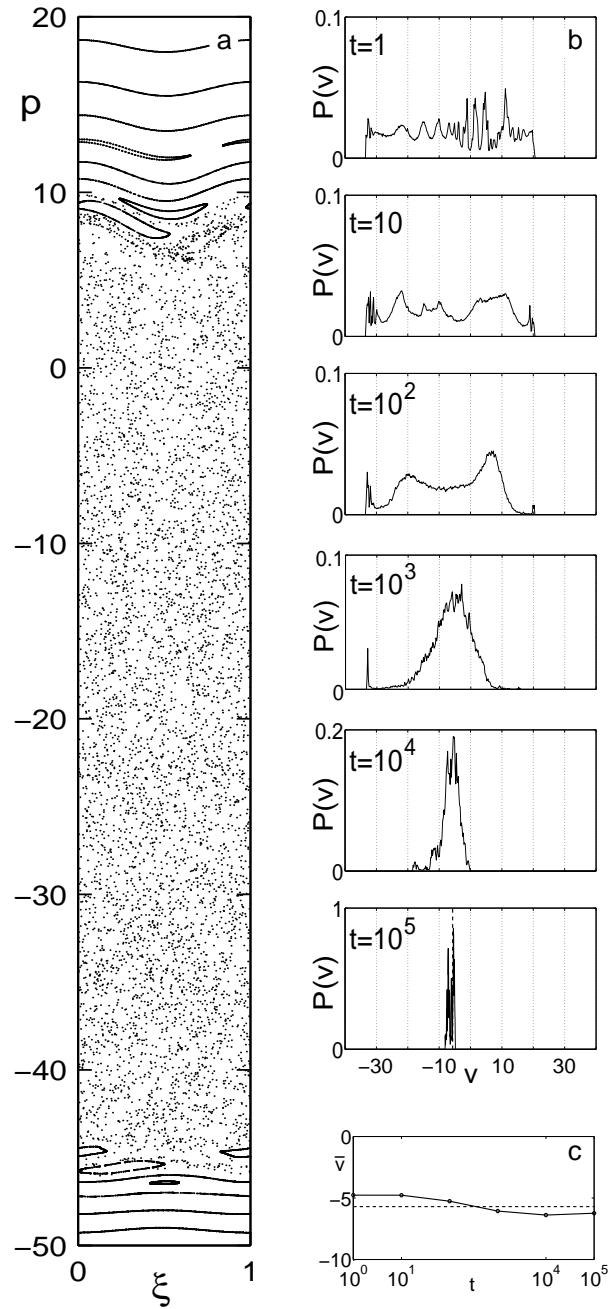,width=8cm}
}
 \caption{\label{fig-ctrw} (a) Stroboscopic Poincar\'e section at $\t=0$
   for the system of Eq.~(\protect\ref{hamnolevy}). (b) For an initial
   distribution $\rho_{0}\sim\chi_{\rm ch}$ the distribution of time-averaged
   velocities is shown at various times $t$. As $t\to\infty$ it evolves to a
   narrow peak around the asymptotic mean velocity (dashed line for $t=10^5$).
   (c) From all distributions shown in (b) the average velocity $\overline v$
   is computed after the contributions from the ballistic channels have been
   removed by restricting $P(v)$ to the interval $-28\le v\le +18$. The
   resulting values (dots) are for all times close to the asymptotic mean
   velocity (dashed).}
\end{figure}

Nevertheless substructures inside the chaotic component of phase space in
general do exist and leave their hallmark in transport properties. L\'evy
walks, in particular, have attracted some attention in the context of
Hamiltonian ratchets \cite{FYZ00,DF01,D+02}. These are the episodes when a
chaotic trajectory is trapped in the vicinity of a transporting island, close
to the hierarchical structure of smaller and smaller islands and surrounding
cantori. Such hierarchical regions are virtually unavoidable in a mixed phase
space (for remarkable exceptions see \cite{Bun01,MP02}). In the context of
ratchets they were termed ``ballistic channels'' \cite{DF01,D+02} and are
frequently located in the vicinity of the KAM tori confining the chaotic sea
from below and above, i.e., in regions of relatively high velocity. Therefore
L\'evy walks are easily observed in numerical transport experiments. Some
care must be taken to avoid the wrong conclusion that ballistic channels and
L\'evy walks are necessary for the existence of substantial chaotic transport
or can completely account for it.

To study this question in some detail, let us start from the sum rule
Eq.~(\ref{SumRule}) and decompose the chaotic transport into contributions
from disjunct subsets of the chaotic sea $C=\bigcup_{j} C_{j}$. We have
$\V_{\rm ch}v_{\rm ch}=\sum_{j}\V_{\rm j}v_{j}$ and $\V_{\rm
  ch}=\sum_{j}\V_{\rm j}$ such that 
\begin{equation}\label{ctrw}
v_{\rm ch}={\sum_{j}\V_{\rm j}v_{j}\over \sum_{j}\V_{\rm j}}\,.
\end{equation}
Because of ergodicity inside the chaotic component the phase-space volumes
$\V_{\rm j}$ in Eq.~(\ref{ctrw}) can be replaced by the fraction of time
a typical chaotic trajectory spends inside subset $j$ or, equivalently, by
the probability to enter subset $j$ times the average survival time in it.
Doing so we immediately arrive at a formula similar in spirit to
Eq.~(3) of Ref.~\cite{DF01} or Eq.~(6) of Ref.~\cite{D+02}. At the same
time it is still exact and does not depend on the character of the subsets
$j$ used to subdivide the chaotic region. As in Refs.~\cite{DF01,D+02},
this decomposition can, e.g., consist of a few prominent ballistic
channels and some remaining chaotic ``bulk'' region. Our main point is
here that in general it is {\em not} possible to approximate this
remainder by an undirected and purely diffusive dynamics, i.e., to set
$v_j=0$ for the corresponding subset in Eq.~(\ref{ctrw}).

For this purpose we will follow the analysis suggested in
Refs.~\cite{DF01,D+02} but apply it to a model with different parameter
values. The Hamiltonian is
\begin{eqnarray}\label{hamnolevy}
H(p,x,t)&=&{p^2\over 2}-2\pi\cos (2\pi x)
\\&&
+(2\pi)^2\,x\,\[2\cos(2\pi t)-4\cos\(4\pi t+{\pi\over 2}\)\]\nonumber
\end{eqnarray}
and the stroboscopic Poincar\'e section (Fig.~\ref{fig-ctrw}a) shows the
typical features discussed in Section~\ref{sec:psec}. The velocity
distribution of the chaotic component is shown in Fig.~\ref{fig-ctrw}b for
various times. In contrast to Fig.~\ref{distributions}c we have chosen here an
ensemble of initial conditions $\rho_{0}\sim\chi_{\rm ch}$ uniformly covering
the entire chaotic sea. Numerically this has been achieved relying on
ergodicity: We run a single long chaotic trajectory $x(t)$ ($0\le t\le 4\cdot
10^5$) and used $x(t')$ with $t'=0,1,2,\dots$ as the initial conditions of the
ensemble. For each such initial condition $\overline v_{t}=[x(t'+t)-x(t')]/t$
is the velocity averaged over a time span $t$. For fixed $t$ the probability
distribution $P(\overline v_{t})$ is shown in Fig.~\ref{fig-ctrw}b. It is
equivalent to the propagator used in Ref.~\cite{D+02} for visualizing internal
details of the chaotic dynamics. Peaks in the propagator can be interpreted as
signatures of partial transport barriers within the chaotic sea. They are
visible as long as the parameter $t$ of the velocity distribution is smaller
than the time scale for crossing the barrier. As expected, for long times
($t>10^{5}$) only a narrow peak survives at a velocity which is in good
agreement with the prediction of the sum rule (dashed line for $t=10^5$).

Since the shape of the velocity distribution depends strongly on time, any
definition of ballistic channels and the corresponding subdivision of the
chaotic invariant set must be highly arbitrary. We single out the most
prominent transporting islands which are visible in Fig.~\ref{fig-ctrw}a close
to the lower and the upper boundary of the chaotic sea. They have winding
numbers $w_{-}\sim 30$ and $w_{+}\sim 20$, respectively. In these regions we
observe particularly sharp peaks in the velocity distribution for $t\lesssim
10^3$ which are signatures of the corresponding L\'evy walks.  Following
Ref.~\cite{D+02} we continue by averaging the velocity distribution over a
region that excludes all such ballistic channels ($-28\le v\le +18$ for the
solid line in Fig.~\ref{fig-ctrw}c; note that this $v$-interval is defined
with respect to the average velocity and therefore is not completely
inside the chaotic layer in the Poincar\'e section shown in
Fig.~\ref{fig-ctrw}a). The result represents the contribution from the
bulk of the chaotic sea. It is definitely non-zero and in fact quite 
close to the asymptotic transport velocity (dashed line), irrespective of the
time scale and the precise cutoff values used. In other words, the average
chaotic transport in this example is mainly due to the bulk region while the
ballistic channels and their L\'evy walks contribute small corrections only. 

This shows that only the invariant sets, as featured in the sum rule
Eq.~(\ref{SumRule}), provide an appropriate concept for the description of the
asymptotic directed transport.


\subsection{Biased ratchets}\label{sec:bias}
\begin{figure}[!b]
 \centerline{\psfig{figure=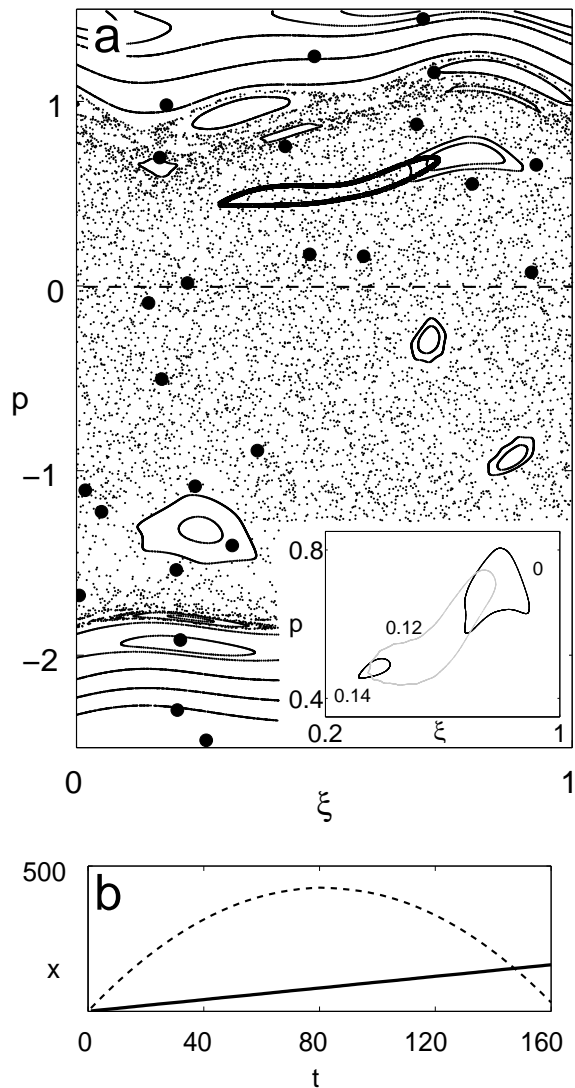,width=8cm}}
 \caption{\label{psec_bias} (a) Stroboscopic Poincar\'e section for the model
   Eq.~(\ref{fyzham}) as in Fig.~\ref{psec_fyz}. On top two trajectories of a
   system with the additional potential $V_{\rm bias}(x)=c\,x$ with $c=0.13$
   are shown.  One trajectory (big dots) was started at $p=10$, i.e., in a
   phase-space region which is in the original system filled by
   non-contractible regular tori. In the presence of the bias such tori are
   absent and the trajectory keeps loosing momentum without bounds. In the
   extended system this trajectory is similar to a parabola (dashed line in
   (b)).  The other trajectory (thick line in (a) and (b)) is part of a
   regular island with winding number $w=1$, i.e., in the extended system this
   trajectory is transporting uphill without loosing momentum. The inset of
   (a) shows the shape of the regular island at different magnitudes of the
   bias potential.  At $c\gtrsim 0.15$ the island disappears. Also islands
   with negative or zero winding number do exist in the biased ratchet (not
   shown).  }
\end{figure}

Can Hamiltonian ratchets be used to transport particles against an external
force? As explained in Sec.~\ref{Hamil}, a constant force does not destroy the
periodicity of the dynamics, and we can still resort to a unit cell to
understand the transport properties. The key question is, which invariant sets
may survive in presence of an additional potential $V_{\rm bias}(x)=cx$.  In
Fig.~\ref{psec_bias}a we compare two trajectories for $c=0.13$ to the familiar
phase-space portrait at $c=0$ (Fig.~\ref{psec_fyz}). One of them was
initialized on a large transporting island with winding number $w=1$. Clearly,
this island is still present although it is distorted and shifted in position.
The winding number of the island is conserved since it is a topological
quantity restricted to rationals. Hence all trajectories inside the islands
have asymptotic mean velocity $\overline v=1$ and we may conclude that
Hamiltonian ratchets can transport uphill! This is confirmed by the full line
in Fig.~\ref{psec_bias}b which shows position vs.~time for the same
trajectory.

The other trajectory was initialized in a phase-space region which for $c=0$
contains non-contractible KAM tori with positive winding numbers. We observe
that for $c=0.13$ the momentum of this trajectory is decreasing without bounds
under the influence of the constant bias force, as naive expectation suggests.
Only in a short time interval, when $p_{t}\approx 0$, the driving potential
has a relevant influence on this trajectory. For long times it behaves
essentially like a free particle accelerated by the bias potential.  Therefore
$x(t)$ for this trajectory is approximately parabolic (dashed line in
Fig.~\ref{psec_bias}b).

From the presence of this single accelerated trajectory we can already
conclude that no regular KAM tori survive in the biased system (at
least not in the phase-space region displayed in
Fig.~\ref{psec_bias}), since these would represent impenetrable
barriers to transport in $p$-direction. Note that the KAM theorem does
not apply to this situation: A constant force does not represent a
smooth perturbation for the unit cell since the potential is not
periodic. In fact there is a simple argument suggesting that an
arbitrary small mean force destroys all non-contractible KAM tori:
Assume that there is a KAM torus of the form $p(\x,\t)$ periodic in
$\x$ and $\t$. Consider its average momentum at some given moment in
time
\begin{equation}\label{meanp}
\overline p(\t)=\int_0^1\d\x\,p(\x,\t)\,.
\end{equation}
As we show by a straightforward calculation in Appendix~\ref{notorus}
the increment of $\overline p$  after one temporal period is given by
\begin{equation}\label{incr}
\overline p(\t+1)-\overline p(\t)=-\int_0^1\d\x\,\d\t\, V'(\x,\t)\,.
\end{equation}
Clearly, this increment must vanish for an invariant KAM torus. However, the
r.h.s.\ of Eq.~(\ref{incr}) is not zero for a biased system with a mean
force. We conclude that no extended KAM tori survive and that therefore the
chaotic sea is no compact invariant set anymore. Hence an arbitrarily small
bias potential will destroy the chaotic ratchet transport in models like
Eq.~(\ref{fyzham}) while uphill transport can be realized by preparing initial
conditions on regular islands of the phase space.


\subsection{A minimal model}\label{sec:minmod}
According to the previous sections, the decisive property of a
Hamiltonian ratchet is an asymmetric mixed phase space. Based on this
insight we can now construct minimal models for Hamiltonian ratchets which
have this property and are otherwise as simple as possible. Probably the
simplest type of model with a mixed phase space are area-preserving maps
generated from kicked one-dimensional Hamiltonians of the form
\begin{equation}\label{kickham}
H(x,p,t)=T(p)+V(x)\sum_{n}\delta(t-n)\,.
\end{equation}
Integrating the equations of motion over one period of the driving we
obtain an explicit map expressing position $x_n$ and momentum $p_n$
immediately before the kick at $t=n$ in terms of the values before the
preceding kick 
\begin{equation}\label{map}
p_{n+1}=p_n-V'(x_{n})\,,
\qquad
x_{n+1}=x_n+T'(p_{n+1})\,.
\end{equation}
The most prominent example is the kicked rotor
\begin{equation}\label{krotorig}
T(p)={p^2\/2}\qquad V(x)={K\over 2\pi}\,\cos(2\pi x)\,,
\end{equation}
one of the best-studied paradigms of Hamiltonian chaos \cite{Izr90}. The phase
space of this model is periodic with period 1 both in $x$ and in
$p$. Therefore one can define a compact unit cell with area $\Delta x\,\Delta
p=1$.

\begin{figure}[!tb]
 \centerline{\psfig{figure=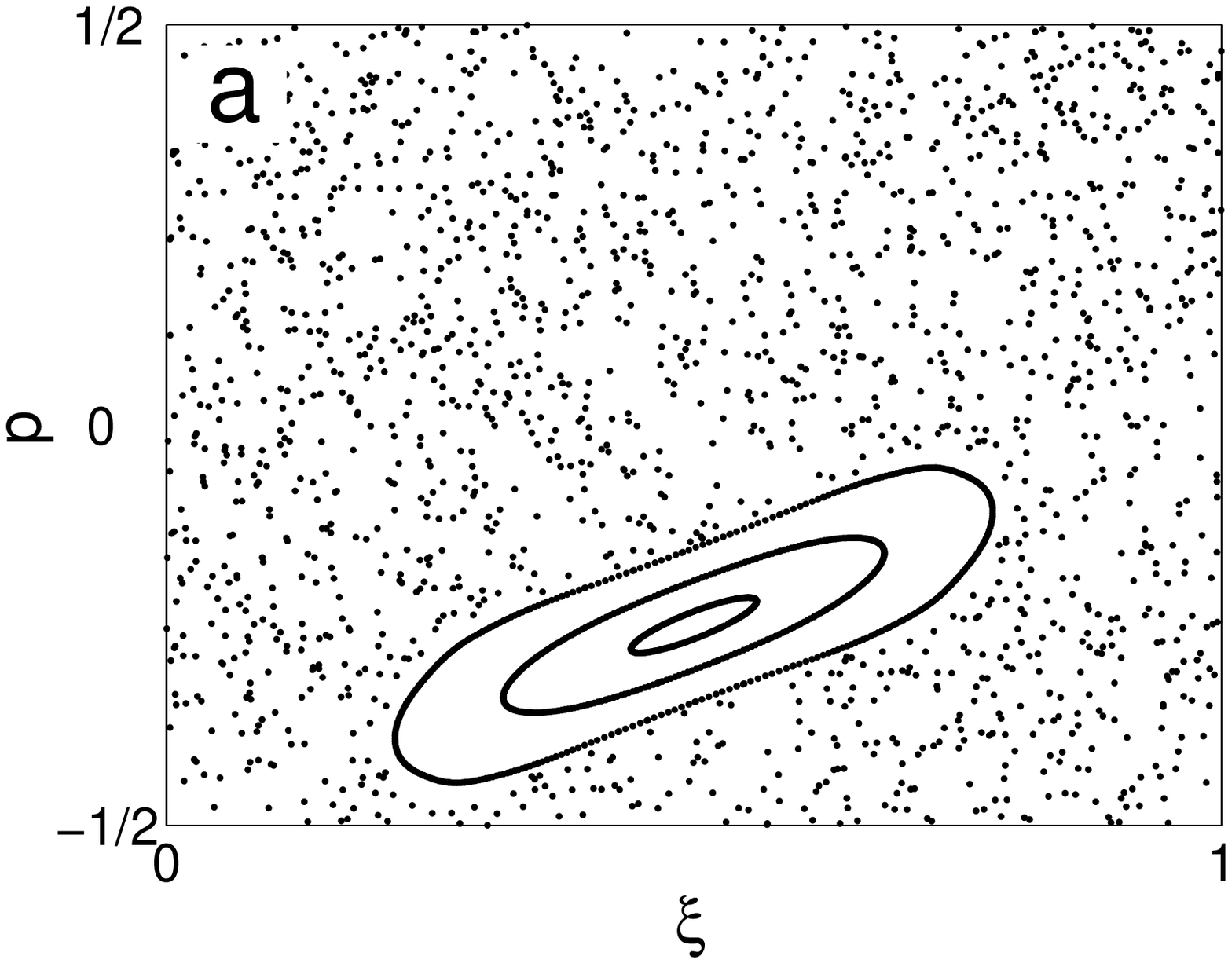,width=7cm}}
 \vspace*{3mm}
 \centerline{\hspace*{5mm}\psfig{figure=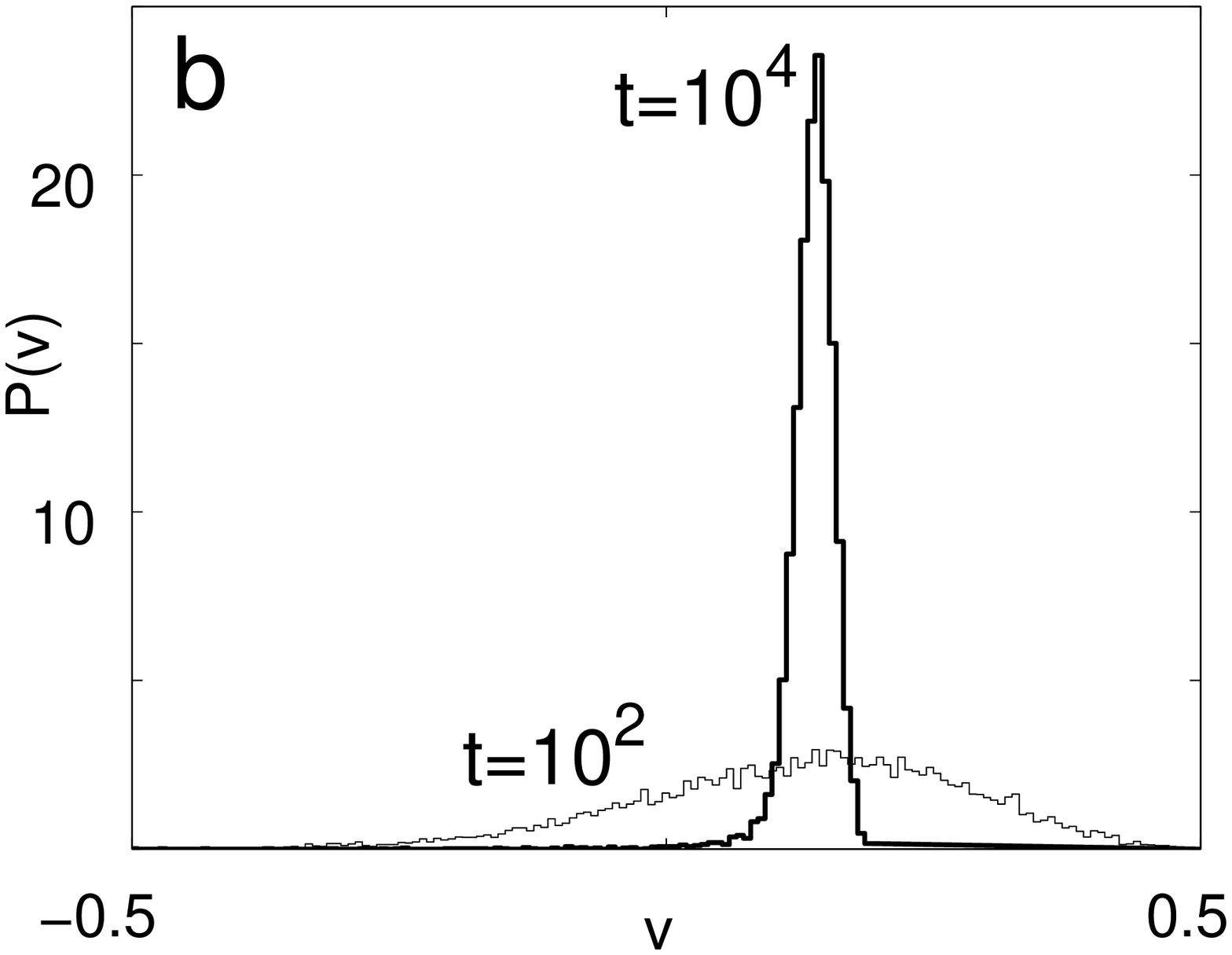,width=7cm}}
 \caption{\label{psecmap} (a) Poincar\'e section $p$ vs $\xi$ of a unit cell for
 the map given by Eq.~(\ref{mfomap}). (b) Velocity distribution $P(v)$ of
 $10^4$ trajectories started at random on the line $p=0$, $x\in [0,1)$ in the
 chaotic sea of the system and iterated until $10^2$ and $10^4$, respectively.}
\end{figure}

The kicked rotor found an important experimental realization in the dynamics
of cold atoms in pulsed laser fields \cite{M+95b,K+98c}. In this experimental
setup the momentum instead of the position is the experimentally accessible
quantity and one is therefore interested in transport along the momentum
direction.  Apart from this purely formal difference, atom optics experiments
promise to be ideal realizations of Hamiltonian ratchets. For this purpose one
has to modify the phase space of the unit cell such that transporting islands
arise and the symmetry $x\to-x$, $p\to-p$ of the kicked rotor is destroyed.

In fact transporting islands appear already in the standard kicked rotor at
kicking strengths $K\gtrsim 2\pi m$. They are referred to as ``accelerator
modes'' \cite{Izr90} and leave traces in the dynamics which were also
experimentally observed \cite{K+98c}. In the kicked rotor these accelerator
modes always come in pairs transporting in opposite directions and therefore
do not lead to transport in the chaotic sea.  However, this symmetry can be
destroyed, e.g., by applying more than a single kick per period or by using
asymmetric potentials in Eq.~(\ref{kickham}).  It is not expected that the
details of these manipulations will be of importance for the resulting chaotic
transport since, as we have shown in the previous sections, the latter is
determined by the underlying phase-space structure only.

In the remainder of this paper we therefore study an abstract model in the
form of Eq.~(\ref{map}). The functions $T(p)$ and $V(x)$ are selected without
reference to any particular experimental setup and only guided by the desire
to have a simple phase-space structure with a large transporting island. We
choose
\begin{eqnarray}\label{mfoham}
V(x)&=&(x\,{\rm mod}\,1\,-1/2)^{2}/2\nonumber\\
T(p)&=&|p|+3\sin(2\pi p)/(4\pi^{2})\,.
\end{eqnarray}
The resulting map
\begin{eqnarray}\label{mfomap}
p_{n+1}&=&p_n-(x_{n}\,\mbox{mod}\,1)+1/2
\nonumber\\
x_{n+1}&=&x_n+\mbox{Sgn}(p_{n+1})+3\cos(2\pi\,p_{n+1})/2\pi
\end{eqnarray}
is considered on a cylinder with transport along the extended $x$-axis while
$p\equiv p+1$ is here a cyclic variable that can be represented with $p\in
[-1/2,+1/2)$.  If the map is restricted to one unit cell
$x\to\x=x\,\mbox{mod}\,1$ we obtain the phase-space portrait shown
in Fig.~\ref{psecmap}a. It shows one large regular island around the stable
fixed point $\x_{0}=1/2$, $p_{0}=-1/4$ with winding number $w_{0}=-1$.
Due to the term $|p|$ in $T(p)$ the phase space has no reflection symmetry
around $p=0$ and also no other momentum-inverting symmetry such that there is
no equivalent island transporting in positive direction. 

There are also no extended regular tori and the whole unit cell must be
considered as the analogue of the compact stochastic layer in the continuously
driven models which we considered in the previous sections.  Consequently the
l.h.s.\ of the sum rule Eq.~(\ref{SumRule2}) vanishes, $0=v_{\rm ch}\,\V_{\rm
ch} + (-1)\,\V_{\rm reg}$. In other words the total transport, averaged over
the whole available phase space, vanishes for this system which confirms that
it is unbiased. A considerable simplification results from the fact that here
the chaotic transport velocity can be computed from the relative phase-space
volume of the single regular island $A_{\rm reg}=1-A_{\rm ch}$ alone,
\begin{equation}\label{krsum}
v_{\rm ch}=A_{\rm reg}/(1-A_{\rm reg})\,.
\end{equation}
From the Poincar\'e section Fig.~\ref{psecmap}a we find $A_{\rm reg}=0.117\pm
0.001$, thus $v_{\rm ch}=0.133\pm 0.001$. This is in very good agreement with
$v_{\rm ch}=0.1344\pm 0.0003$ obtained directly from the spatial
distribution of $10^4$ trajectories after $2\times 10^4$ kicks.
Fig.~\ref{psecmap}b shows the convergence of the chaotic velocity distribution
to a delta function concentrated at this value, in accordance with
Eq.~(\ref{nospread}).

We would like to stress again that the directed chaotic transport in this
ratchet model is a consequence of the phase-space structure and cannot be
explained by the asymmetric kinetic-energy function alone. We have verified
this fact by repeating the analysis for a larger potential $5V(x)$. Then the
phase space is completely chaotic, yet despite the asymmetric function $T(p)$
no transport is observed.


\section{Quantum Ratchets}

We now turn to the investigation of quantized Hamiltonian ratchets, i.e.,
driven 1D Hamiltonian quantum systems which are classically periodic both in
space and in time. We restrict attention to systems in which the phase-space 
volume of a unit cell is finite and phase space is composed of a
chaotic sea with one or more embedded regular islands, as in the minimal
ratchet model discussed above. This restriction leads to a finite 
Hilbert-space dimension which simplifies the numerical calculations. Moreover,
we have seen that the dynamical processes relevant for transport are
restricted anyway to the compact chaotic layer of the unit cell. We
therefore expect models with finite Hilbert-space dimension to capture
also the essential features of quantized ratchet transport.

\subsection{Floquet operator and eigenstates}\label{sec:floqop}

For a system periodic in time, one can still construct a dynamical group with
a single time-like parameter, which however now becomes discrete, measuring
time in units of the period of the driving. It is generated by the unitary
evolution operator over one period,
\begin{equation}\label{floqop}
\hat U(t+1,t)= \hat\T \exp\left(-\frac{\i}{\hbar}\int_0^1\d t\,
\hat H(t)\right),
\end{equation}
where $\hat\T$ effects time ordering. The computation of this Floquet
operator is simplified considerably if $H(t)$ is a kicked Hamiltonian
as in Eq.~(\ref{kickham}). Then the time evolution from time
$t=m-\varepsilon$ immediately before the kick $m$ to time
$t=m+1-\varepsilon$ immediately before the following kick can be expressed in
terms of $T(p)$ and $V(x)$ as a product
\begin{equation}\label{floqopkick}
\hat U=\e^{-\i T(\hat p)/\hbar}\e^{-\i V(\hat x)/\hbar}
\end{equation}
of two operators which are diagonal in the position or the momentum
representation, respectively. The time evolution of a state is obtained by
successive multiplications by phase factors and fast Fourier transforms
effecting a basis change. An additional simplification results if we consider
$p$ as a cyclic variable $p\equiv p+1$, as is the case with the minimal
ratchet model Eq.~(\ref{mfomap}) to which our numerical results will be
restricted. In this case the wave function is periodic in $p$ with
$\psi(p+1)=\psi(p)$ and consequently the conjugate variable $x$ is restricted
to the discrete values $x_{n}=nh$.  Here, $h$ denotes the dimensionless ratio
of Planck's constant to the phase-space area of the classical unit cell which
we set to unity in Eq.~(\ref{mfomap}). It is a well-known peculiarity of
models with this property that the periodicity of the classical potential
$V(x+1)=V(x)$ (or at least $V'(x+1)=V'(x)$ in the case of our minimal model)
does not necessarily lead to a spatially periodic Floquet operator.  The
reason is that the potential is now restricted to discrete values
$V_{n}=V(x_{n})=V(nh)$ and periodicity is achieved only if there is an integer
$N$ with $V_{n+N}=V_{n}$ which implies $hN=M$ with another integer $M$. Hence
$h=M/N$ must be rational. In contrast, in periodic systems with infinite
phase-space volume such as Eq.~(\ref{fyzham}), the Floquet operator is
spatially periodic irrespective of the value of Planck's constant. In the
following sections we shall use values $h=1/N$ to ensure that the
quantum system has the same spatial periodicity as the classical model. Only
in the last Section~\ref{sec:disorder} we consider modifications of our
results for irrational values of $h$. They are to be interpreted as a spatial
disorder that does not affect the classical phase-space structure but destroys
the perfect periodicity of the corresponding quantum system.

A double periodicity, both in space and time, requires to combine the
corresponding representations of quantum mechanics appropriate for these
symmetries, i.e., Bloch and Floquet theory, respectively. The
eigenvalue equation 
\begin{equation}
|\phi_{\alpha}(t+1)\>=
\hat U|\phi_{\alpha}(t)\>=\e^{-2\pi
\i\epsilon_{\alpha}}\,|\phi_{\alpha}(t)\>
\end{equation}
defines \emph{Floquet states} $|\phi_{\alpha}\>$ and
\emph{quasienergies} $\epsilon_{\alpha}\in [0,1)$ \cite{Sam73}. For the
systems considered here $\alpha$ is a discrete index $1\le\alpha\le N$.  

\begin{figure}[htb]
 \centerline{\psfig{figure=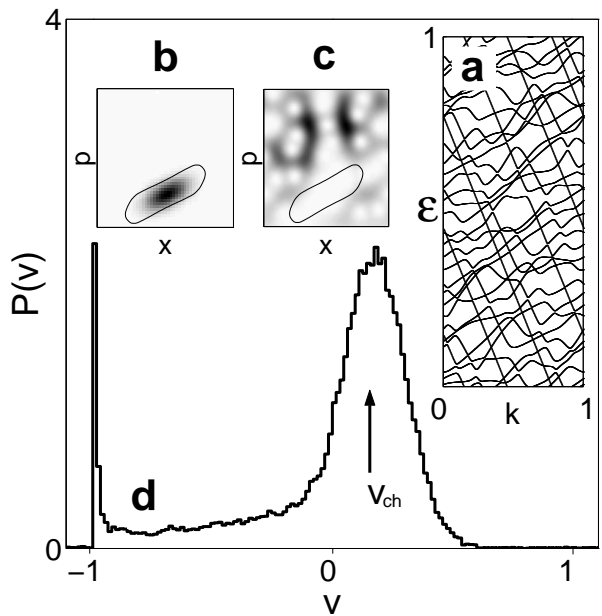,width=8cm}}
  \caption{\label{fig:velo} (a) Quasienergy band spectrum of the minimal
    ratchet model Eq.~(\protect\ref{mfomap}) at $h^{-1}=32$. Regular bands
    appear as approximately straight lines with negative slope. (b) The Husimi
    representation of the Floquet eigenstates corresponding to points on these
    lines are concentrated inside the regular island. (c) Most other
    eigenfunctions spread over the entire chaotic sea but avoid the regular
    island. The corresponding bands have strongly fluctuating slopes. (d)
    Distribution of band slopes (velocity expectation values) at $h^{-1}=128$.
    The sharp peak at $v=-1$ corresponds to the regular bands, the broader
    peak to the chaotic bands. The velocity of the classically chaotic
    transport is marked by an arrow.}
\end{figure}

For the discrete spatial translation group there is a
continuous set of representations parameterized by the
quasimomentum $k\in[0,1)$. In the simultaneous presence of temporal
periodicity, the Bloch theorem now applies to Floquet states,
\begin{equation}
\phi_{\alpha,k}(x+1,t)=\e^{2\pi\i k}\phi_{\alpha,k}(x,t)
\end{equation}
so that both eigenstates and eigenphases carry a double index $(\alpha,k)$.
The support of the \emph{Floquet band spectrum}, in all cases considered here,
consists of continuous lines in the two-dimensional $(k,\epsilon)$-space, cf.\ 
Fig.~\ref{fig:velo}a.  Since the spectrum is periodic with period 1 both in
$\epsilon$ and $k$, these variables are canonically conjugate to a pair of
integers $(\nx,\mt)$ which measure position and time in units of the spatial
and temporal periods, respectively. We shall show in Section
\ref{sec:shorttime} that this allows to relate the band structure to the time
evolution of the spatial distribution by a double Fourier transformation.

We have seen that the decisive property of classical Hamiltonian ratchets is
the existence of invariant sets of phase space with different average
velocities. Traces of the classically invariant sets are manifest in the
quantum dynamics only if the quantum uncertainty allows to resolve them, i.e.,
if $\hbar$ is much smaller than the relevant phase-space structures. From here
on we shall restrict our attention to this semiclassical regime. 
Fig.~\ref{fig:velo}a shows an example of a Floquet band spectrum for a
Hamiltonian ratchet, the minimal model (\ref{mfomap}). 
This system has two distinct invariant sets in phase space, the 
chaotic sea and one transporting island embedded in it. According
to the semiclassical eigenfunction hypothesis \cite{Per73,Ber77} one expects
that in the semiclassical regime almost all eigenfunctions condense on one 
of the invariant phase-space sets. Fig.~\ref{fig:velo}b, c 
show the Husimi representations \cite{H+84} of typical eigenstates. 
Indeed, one of them is concentrated inside the regular island while the other
populates the chaotic sea, avoiding the island. Associated with these two
types of eigenstates are two types of bands: regular bands appear in the
spectrum as straight lines with slope $\d\epsilon_\alpha/\d k\approx -1$,
chaotic bands are fluctuating and have on average a positive slope. 
In the subsequent section, we are going to make this relation between
bands and subsets of phase space more precise. We use it to establish
a sum rule for transport in quantum ratchets analogous to the
classical sum rule discussed in Section~\ref{sec:clsumrule}.

\subsection{Semiclassical transport in terms of Floquet bands}
\label{sec:shorttime}
\subsubsection{Quantum sum rule}

The basic relation expressing the velocity of a Floquet state
in terms of the quasienergy band to which it belongs is
\begin{equation}\label{velo}
v_{\alpha,k}=\<\<{\phi_{\alpha,k}|\hat v|\phi_{\alpha,k}}\>\>=
{{\rm d}\epsilon_{\alpha,k}\/{\rm d}k}\,.
\end{equation}
In the present case of a periodically driven system, the expectation
value of the velocity operator $\hat v=\hat T'(\hat p)$ includes a
time average over one period of the driving $\langle\langle \ldots
\rangle\rangle \equiv \int_{0}^{1}{\rm d}t\<\ldots\>$. The second
member of Eq.~(\ref{velo}) then follows from applying the
Hellman-Feynman theorem, which was proven for time-periodic systems in
\cite{Sam73}.

A wave packet localized on the scale of a single unit cell or
narrower corresponds to a nearly homogeneous
distribution in $k$. The corresponding mean velocity for a whole band
$\alpha$ vanishes, 
\begin{equation}
\label{velok}
\< v_{\alpha}\>_{k}=\int_0^1{\rm d}k\,{{\rm d}\epsilon_{\alpha,k}\/{\rm
d}k}=0\,,
\end{equation}
as is implied by the periodicity of the bands. Averaging also over
energy, i.e., summing over the bands, we find as velocity average over
the total Hilbert space of the unit cell,
\begin{equation}
\label{qsumrule}
\< v\>_{k,\epsilon} = {1\/N}\sum_{\alpha} \l<{{\rm d}\epsilon_{\alpha,k}\/{\rm
d}k}\r>_{k} = 0\,.
\end{equation}
Equation (\ref{qsumrule}) can be considered the quantum-mechanical
counterpart of the classical sum rule for transport, Eq.~(\ref{SumRule2}).
Effectively, the quantum sum rule like the classical one refers to a 
finite subset of phase space. Here, the cutoff is introduced by the
finite dimension of the basis used to span the Hilbert space of the
unit cell in calculating the band spectrum. 

The crucial step for this quantum sum rule is the averaging along a given band
$\alpha$ over the entire Brillouin zone, Eq.~(\ref{velok}).  In particular,
this amounts to regarding all band crossings, however narrow, as avoided
crossings. If $k$ were considered a parameter with a fictitious time
dependence, the quantum time evolution under a slow change of $k$ would
respect avoided crossings in exactly this manner. Therefore these bands are
referred to as \emph{adiabatic bands} \cite{KMG94}.

It follows, conversely, that a finite mean velocity can be
obtained if modified bands are constructed by connecting band segments 
\emph{across} all avoided crossings with a gap below some threshold. 
Such bands determine the time evolution under a fast change of $k$ and
accordingly are called \emph{diabatic} \cite{KMG94}. They are not
associated to a fixed band index $\alpha$ and therefore need not be
periodic in $k$. So for individual diabatic bands Eq.~(\ref{velok}) does
not apply, their mean velocity can be finite. We argue in the following
that indeed it is diabatic bands, not adiabatic ones, which semiclassically
correspond to invariant sets of classical phase space, and to which
a relation between band structure and directed transport must refer.

Fig.~\ref{fig:velo} provides numerical evidence to justify the assignment of
invariant sets to diabatic bands.  For example, the regular island with
winding number $-1$ is associated to straight-line segments in the spectrum,
corresponding to a quantum velocity $v_{\alpha,k}=-1$ with very small
fluctuations. In contrast, chaotic regions are
represented by ``wavy'' band sections with strongly varying slope to which a
precise velocity value cannot be assigned. In this sense, it is legitimate to
talk of ``regular'' vs.\ ``chaotic'' diabatic bands.

In the following we will reconsider the sum rule Eq.~(\ref{qsumrule}) using
diabatic bands and express the different contributions in terms of the
invariant sets of the classical phase space.  First we note that replacing in
Eq.~(\ref{qsumrule}) adiabatic by diabatic bands amounts to interchanging band
indices at avoided crossings, thus it results at most in a permutation of
terms within the sum but does not affect the sum rule as a whole. We can
therefore group diabatic-band terms in Eq.~(\ref{qsumrule}) according to the
classical invariant set they pertain to,
\begin{equation}\label{sumregchb}
0=\sum_{\alpha\;\in\;\rm ch.\; bands} 
\l<{{\rm d}\epsilon_{\alpha,k}\/{\rm d}k}\r>_{k}
+
\sum_{r\;\in\;\rm reg.\; bands} \l<{{\rm d}\epsilon_{r,k}\/{\rm d}k}\r>_{k}.
\end{equation}
In the semiclassical limit the respective numbers of terms in the sums
are given by the relative fraction of phase space occupied by the
corresponding invariant sets, i.e., $N_{\rm ch}=f_{\rm ch}\,N$ for the
chaotic bands and $N_{\rm r}=f_{\rm r}\,N$ for the various embedded
regular islands $r$. $N=h^{-1}$ is here the total number of bands,
i.e., the Hilbert-space dimension per unit cell. Assuming that
the classical phase space contains only a single chaotic component we
can characterize the associated diabatic bands by a mean slope
$\<v_{\rm ch}\>$ and have $\sum_{\alpha\in{\rm ch}}\<{{\rm
d}\epsilon_{\alpha,k}/{\rm d}k}\>_{k}= N_{\rm ch}\<v_{\rm ch}\>$. 

For the regular bands, the double periodicity of the $(k,\epsilon)$-space
allows to define \emph{winding numbers} in the same way as we did in Section
\ref{sec:psec} for the topology of regular islands in the conjugate
$(x,t)$-space. For the same reason as the classical winding numbers these
topological quantum numbers have to be rational, i.e., $w^{\rm qm} = n/m$ if
the band closes upon itself after $n$ revolutions in $\epsilon$- and $m$
revolutions in $k$-direction. As the regular states are localized on the
invariant tori inside the island, their velocity expectation (band slope)
in the semiclassical limit approaches the regular transport velocity.
This leads to the conclusion
\begin{equation}
\label{qwinding}
\l<{{\rm d}\epsilon_{r}\/{\rm d}k}\r>_{k}\approx w_{r}^{\rm qm}=
w_{r}^{\rm cl} = v_{r}^{\rm cl}\,.
\end{equation} 
Avoided crossings modify the band slopes in a range which is negligible in the
semiclassical limit (see Sec.~\ref{sec:longtime}), while the winding numbers
as topological quantities are not affected at all. In other words, the winding
number $w^{\rm qm}_r$ of a diabatic band $r$ pertaining to a classical regular
island $r$ is {\em identical} to the classical winding number $w^{\rm cl}_r$
of that island, Eq.~(\ref{wz}).  We have now
\begin{equation}
0 = N\,f_{\rm ch} \<v_{\rm ch}\> +
N\sum_{r} f_{r}v_{r}^{\rm cl}\,.
\end{equation}
Note that $f_{\rm ch}$, $f_{r}$, and $v_{r}^{\rm cl}$ are all \emph{classical}
quantities. Consequently, also the quantum transport velocity $\<v_{\rm ch}\>$
must coincide with its classical counterpart
\begin{equation}\label{qvch}
\<v_{\rm ch}\>=v_{\rm ch}^{\rm cl}\,. 
\end{equation}
This is the main result of the quantum mechanical sum rule. We stress again
that it pertains to the semiclassical regime since otherwise the notion of
diabatic bands is not applicable.

Fig.~\ref{fig:velo}d confirms Eq.~(\ref{qvch}) qualitatively. It shows the
distribution of quantum velocities (band slopes) for our minimal ratchet
model. We observe two well separated peaks, one for the regular bands at
$v_r^{\rm cl}=-1$ and one at $v_{\rm ch}^{\rm cl}$ for the chaotic bands.  The
region separating the two peaks corresponds to the band slopes in the vicinity
of avoided crossings between regular and chaotic bands.  The weight of the
distribution in this intermediate region decreases with $h$ and vanishes in
the semiclassical limit $h\to 0$.


\subsubsection{Form factor}
\def\wt{\mu} \def\wx{\nu} 
Our analysis based on winding numbers can be applied to predict the
mean quantum transport velocity in the semiclassical regime from the
classical value. The band spectra, however, contain more detailed
information about quantum transport, encoded in the spectral two-point
correlation functions. A double Fourier transform $\epsilon \to {\mt}$,
$k \to {\nx}$ and subsequent squaring of the spectral density translates
two-point correlations in the bands into the entire time evolution of the
spatial distribution on the scale of the temporal and spatial periods,
respectively. 

As a suitable quantity to establish this relation, we recur to the
generalized form factor introduced and studied in \cite{D+98} for
completely chaotic systems. We define it as
\begin{equation}\label{ff}
K({\nx},{\mt})={1\over N}\langle|u({\nx},{\mt})|^2\rangle
\end{equation}
with
\begin{eqnarray}\label{spamp}
u({\nx},{\mt})&=&\int_{0}^{1}{\rm d}k\,\e^{2\pi\i k{\nx}}\Tr U_{k}^{{\mt}}
\nonumber\\&=&
\sum_{\alpha=1}^{N}\,\int_{0}^{1}{\rm d}k\,
\e^{2\pi\i (k{\nx}-\epsilon_{\alpha,k}\mt)}
\nonumber\\&=&\sum_{\alpha=1}^{N} u_{\alpha}({\nx},{\mt})\,.
\end{eqnarray}
$N$ denotes the Hilbert-space dimension per unit cell, that is the
phase-space area of a unit cell in units of Planck's constant
$h$. $U_{k}$ is the $N\times N$ Floquet operator (\ref{floqop})
evaluated at Bloch number $k$. The integers ${\nx}$, ${\mt}$ are the
discrete variables canonically conjugate to $k$ and $\epsilon$,
respectively, that is, the unit-cell number relative to the starting
point, and time in units of the period of the driving. The average
$\langle\ldots\rangle$ in Eq.~(\ref{ff}) is essential in order to
remove the otherwise dominant fluctuations around the mean value. It
can be taken over a narrow time range or over an ensemble of quantum
systems corresponding to approximately the same classical system. 

As we will now show, the form factor is related, on the one hand, to the
classical dynamics of a distribution which initially covers homogeneously
the phase space of a single unit cell. On the other hand, it contains the
quantum velocity distribution as a limiting case. Therefore it is an
appropriate starting point for a semiclassical theory of ratchet transport.

We assume to be sufficiently close to the semiclassical limit $N\gg 1$ such
that we can consider the band spectrum in the diabatic approximation.
Moreover, in the semiclassical limit it is justified to neglect correlations
between diabatic bands pertaining to different invariant sets (regular or
chaotic) unless they are related by symmetries. This allows to write the form
factor as an incoherent sum of the respective contributions, because the
averaging in Eq.~(\ref{ff}) suppresses uncorrelated cross terms. We obtain
\begin{equation}
K({\nx},{\mt}) = \sum_r K_{r}({\nx},{\mt}) + K\ch({\nx},{\mt})\,,
\end{equation}
the sum running over all regular invariant sets (islands and island
chains). 

In Appendix \ref{app:ffreg} we obtain the semiclassical expression 
\begin{equation}\label{kr}
K_{r}({\nx},{\mt}) =
f_r\,\wt_r\delta_{\wt_r{\nx}-\wx_r{\mt}}
\,.
\end{equation}
for the form factor of a chain of regular islands with winding number
$w_{r}=\wx_r/\wt_r$. It seems that the form factor is enhanced by a factor
$\wt_r$ for an island chain as compared to a single island of equal total
size, but this is not the case.  In Eq.~(\ref{kr}),
$\delta_{\wt_r{\nx}-\wx_r{\mt}}=1$ holds only at the unit cell
${\nx}=(\wx_r/\wt_r){\mt}=v_{r}\mt$ to which a classical trajectory, started
in the regular island at ${\nx}=0$, has traveled in time ${\mt}$. In
particular, as $\nx$ is an integer, ${\mt}$ must be an integer multiple of
$\wt_r$. That is, $K_{r}({\nx},{\mt})$ is finite only every $\wt_r$-th period
of the driving, such that the average contribution to the form factor is
independent of the period  $\wt_r$ of the island chain. 

For the chaotic contribution to the form factor we can resort to a
semiclassical theory which has been developed for completely chaotic
systems in \cite{D+97,D+98}. In order to apply it to a system with a mixed
classical phase space we assume the validity of the ergodic sum rule
\cite{HO84} for the chaotic component. Then the
result of \cite{D+97,D+98} remains essentially unchanged, and the form
factor is given in terms of the classical velocity distribution of the
chaotic component as 
\begin{equation}\label{ksc}
K\ch({\nx},{\mt}) = {{\mt}\/m_{\rm H}}\,P\ch\({{\nx}\/{\mt}},{\mt}\)\qquad
(\mt\lesssim m_{\rm H})\,.
\end{equation}
To be precise, $P_{\rm ch}(v,t)$ entering this equation is the chaotic
classical propagator for a uniform distribution inside the chaotic sea, as
introduced in Section \ref{sec:barriers}.  Its definition is
Eq.~(\ref{velodist}) with $\rho_0=\chi_{\rm ch}$.  Since Eq.~(\ref{ksc}) is
based on the diagonal approximation \cite{Ber85}, i.e., correlations between
different classical orbits have been neglected, it is valid only for short
times and breaks down beyond the Heisenberg time $m_{\rm H}\approx N\ch\approx
f\ch N$ of the chaotic component.

\subsubsection{Quantum velocity distribution}

A complementary approximation to the form factor for long times can be
achieved following again Refs.~\cite{D+97,D+98}. The chaotic bands fluctuate
as a function of $k$ with an amplitude approximately given by the spacing
$\Delta\epsilon \approx N\ch^{-1}$ between them. For times beyond the
Heisenberg time, these fluctuations give rise to phase oscillations in the
integrand of Eq.~(\ref{spamp}) which exceed $2\pi$. Therefore we can perform
the $k$-integration in stationary-phase approximation and obtain
\begin{eqnarray}
u_{\alpha}({\nx},{\mt})&=&\sum_{\epsilon_{\alpha,k_{\rm s}}'={{\nx}/{\mt}}}
\sqrt{\i/|\epsilon_{\alpha,k_{\rm s}}''{\mt}|}\nonumber\\
&&\times\exp(2\pi\i [k_{s}{\nx}-\epsilon_{\alpha,k_{\rm s}}{\mt}])\,,
\end{eqnarray}
i.e., only those points $k=k_{\rm s}$ contribute to the integral 
where the derivative of the phase of the integrand vanishes,
$0={\nx}-\epsilon_{\alpha,k_{\rm s}}'{\mt}$. These are isolated
points in the spectrum which can be assumed to vary independently upon
averaging in Eq.~(\ref{ff}). Therefore we can neglect all cross terms when
squaring the sum of contributions from different points of stationary
phase and obtain for the form factor
\begin{eqnarray}\label{kpsp}
K\ch({\nx},{\mt})&=&
\frac{1}{m_{\rm H}\,{\mt}}
\sum_{\alpha}\sum_{\epsilon_{\alpha,k_{\rm s}}'={{\nx}/{\mt}}}
|\epsilon_{\alpha,k_{\rm s}}''|^{-1}\!\!.
\end{eqnarray}
Now that we are rid of all phase factors it is very instructive to
rewrite the result again as an integral over the Bloch number $k$ 
\begin{equation}\label{kintk}
K\ch({\nx},{\mt})=
\frac{1}{m_{\rm H}\,{\mt}}
\sum_{\alpha}\int_{0}^{1}\d k\,\delta\(\epsilon_{\alpha,k}'-{{\nx}\/{\mt}}\)\,.
\end{equation}
This equation has two important consequences. First we note that up to
normalization the form factor beyond the Heisenberg time is nothing but the
distribution of band slopes alias quantum velocities 
\begin{equation}\label{ffvelodist}
K(\nx,\mt)\sim P_{\rm quant}(v)|_{v=\nx/\mt}
\qquad(\mt>m_{\rm H})\,, 
\end{equation}
which is shown for the minimal model in Fig.~\ref{fig:velo}a.  As in the
classical case, this velocity distribution is the natural quantity to describe
a system with directed ballistic quantum transport and the form factor can be
considered a useful generalization of it.  

Secondly, Eq.~(\ref{kintk}) implies that the form
factor at any time ${\mt}$ beyond the Heisenberg time $m_{\rm H}$ can be
expressed via scaling by the form factor right at the Heisenberg time
\begin{eqnarray}\label{kball}
K\ch({\nx},{\mt})&=&{m_{\rm H}\/{\mt}}\,
K\ch\(m_{\rm H}\,{{\nx}\/{\mt}},m_{\rm H}\)
\nonumber\\&=&
{m_{\rm H}\/{\mt}}\,
P_{\rm ch}\({{\nx}\/{\mt}},m_{\rm H}\)\ (\mt>m_{\rm H})\,.
\end{eqnarray}
In the second line we have used the semiclassical approximation
Eq.~(\ref{ksc}) for $\mt=m_{\rm H}$. It is valid only up to the Heisenberg
time, but according to Eq.~(\ref{kintk}) it determines the form factor also
beyond. Of course, the validity of Eq.~(\ref{kball}) depends on applying both,
the short-time and the long-time approximations for the form factor right at
the Heisenberg time where they are on the verge of breaking down.  This
interpolation procedure has been corroborated by comparison to results from
numerics and from supersymmetry in \cite{D+97,D+98}. We expect that it applies
in the present case of a transporting chaotic component as well.

The two consequences of Eq.~(\ref{kintk}) combine to the conclusion that the
distribution of quantum velocities in the chaotic component of the band
spectrum is equal to the distribution of time-averaged classical velocities
for an ensemble of particles filling the chaotic component of phase space
homogeneously. Information on the quantum system enters into this classical
distribution only via the point in time at which this velocity distribution is
evaluated---it must be chosen as the Heisenberg time $N\ch$ of the chaotic
component. Before writing down this result we note that the restriction to the
chaotic component is actually not necessary, since for the embedded regular
islands the same result applies trivially because of Eq.~(\ref{qwinding}).
Hence we have 
\begin{equation}\label{qvd}
P_{\rm quant}(v)=P_{\rm class}(v,m_{\rm H})\,,
\end{equation}
for a stochastic layer including one chaotic component and one or more
embedded regular islands. Eq.~(\ref{qvd}) is a nontrivial result because it
establishes quantum-classical correspondence for the velocity distributions
and thus for asymptotic {\em long-time} transport properties. We stress again
that this result was derived semiclassically within the diagonal
approximation. It would be very interesting to explore possible corrections
due to neglected interferences between classical periodic orbits (akin the
weak-localization correction in the standard form factor \cite{Ber85}), but at
present the methods to deal with such corrections \cite{M+04} are not
sufficiently developed to treat the type of system we are dealing with here.

\subsection{Long-time quantum transport and dynamical tunneling}
\label{sec:longtime}
\subsubsection{Transport of wave packets}

So far we have considered transport only in terms of stationary quantities
like eigenstates and band spectrum. Using the obtained results we can now
describe the transport of arbitrary wave packets.  The asymptotic quantum
transport velocity of a wave packet is an average over all band slopes,
weighting each Floquet state by its overlap with the initial state. To see
this we write the wave packet as a superposition of Floquet states
\begin{eqnarray}\label{tdwp}
\psi(x,t)&=&
\sum_{\alpha}\int_{0}^{1}{\rm d}k\,
 \psi_{\alpha,k}(t)\,\phi_{\alpha,k}(x)
\nonumber\\&=&
\sum_{\alpha}\int_{0}^{1}{\rm d}k\,
 \psi_{\alpha,k}\,\e^{-2\pi\i\,\epsilon_{\alpha,k}t}\,\phi_{\alpha,k}(x)\,,
\end{eqnarray}
calculate the expectation value $\< x(t)\>$ of position as a function of
time (see App.~\ref{app:wt}), and obtain
\begin{eqnarray}
\< x(t)\>&=&\int_{-\infty}^{+\infty}{\rm d}x\,x|\psi(x,t)|^2
\nonumber\\&=&
v_{\infty}t+{\rm o}(t)\,,
\label{wptrans}
\end{eqnarray}
with
\begin{equation}
\label{wpvel}
v_{\infty}=\int_{0}^{1}{\rm d}k\,\sum_{\alpha}
|\psi_{\alpha,k}|^{2}\,v_{\alpha,k}\,.
\end{equation}
\begin{figure}[htb]
 \def\fw{40mm} \centerline{
  \psfig{figure=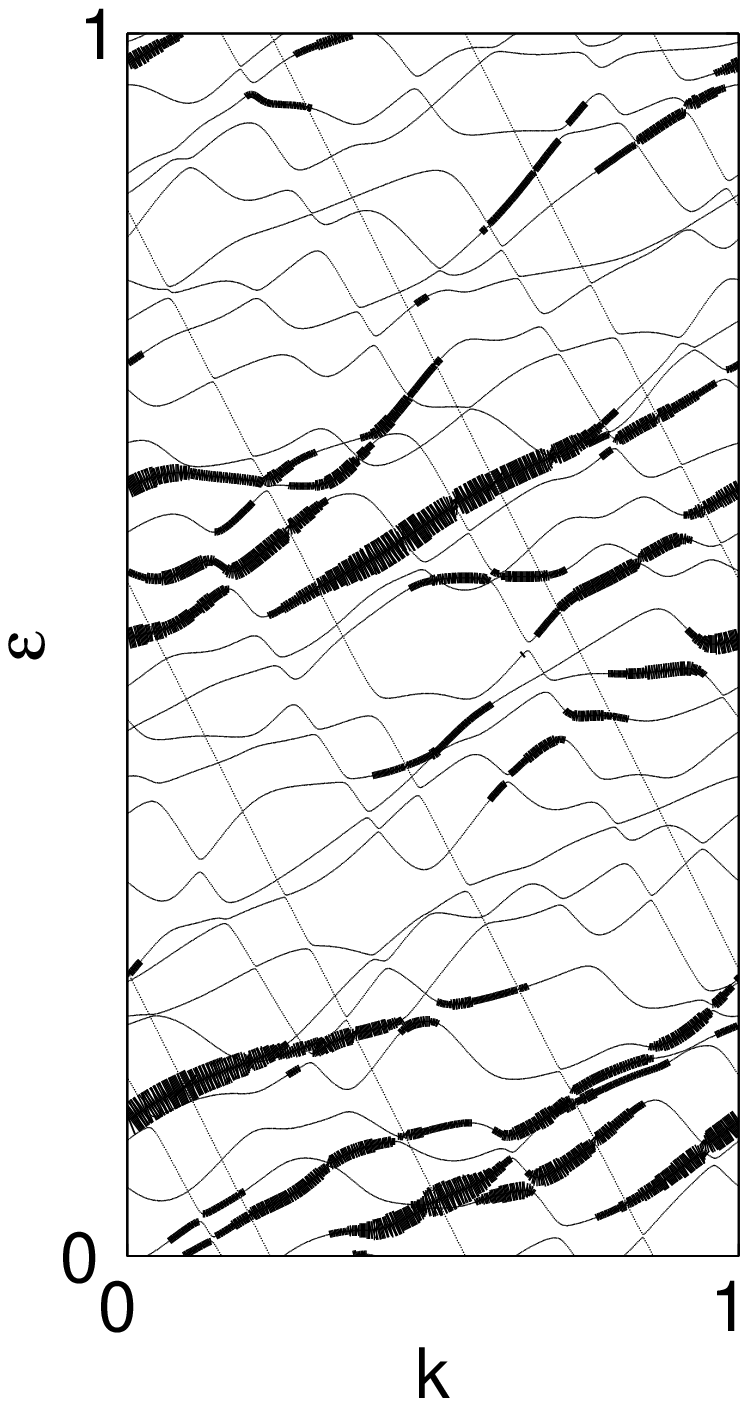,width=\fw}
  \psfig{figure=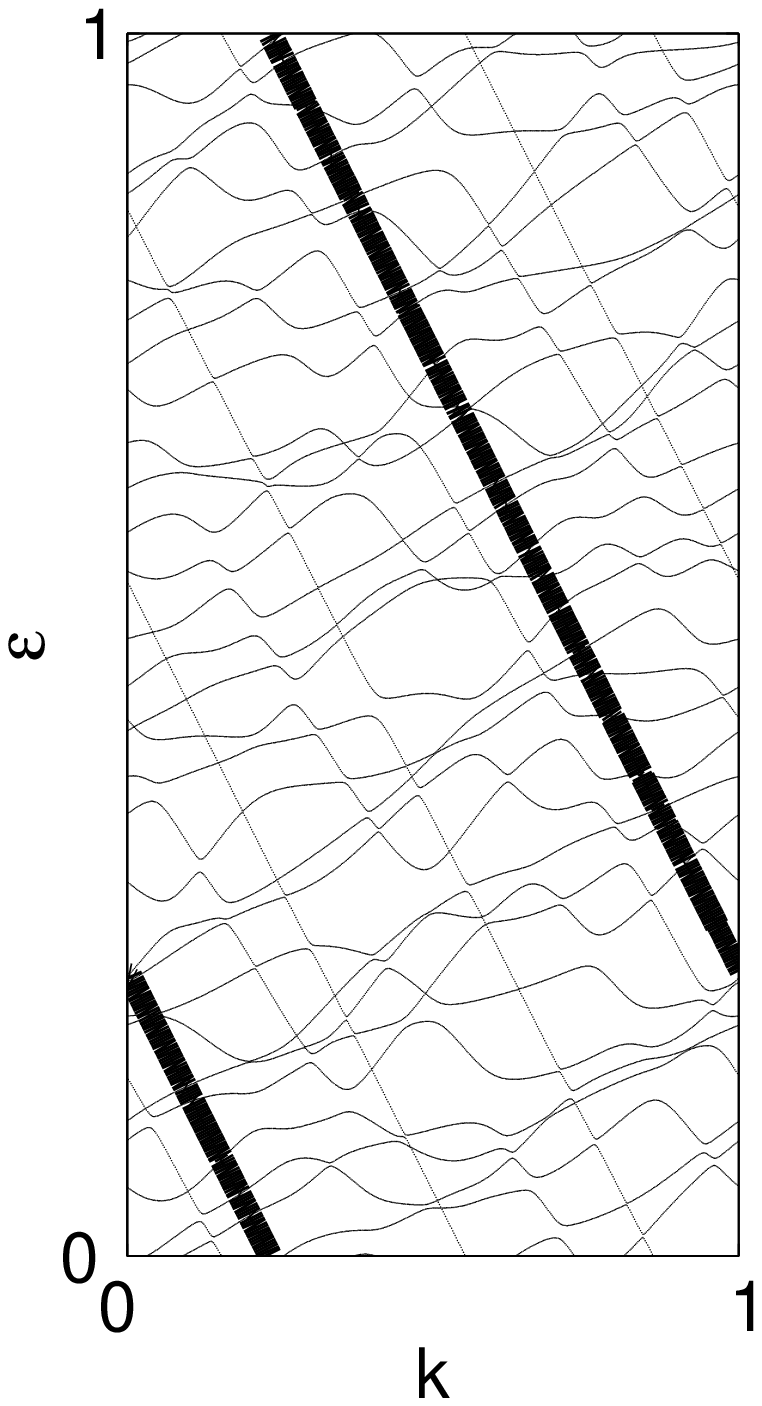,width=\fw}
 }
 \caption{\label{fig:bw} Quasienergy band spectrum of the minimal ratchet
 model at $h^{-1}=32$. The linewidth encodes the overlap
 $|\<\phi_{\alpha,k}|\psi\>|^{2}$ of the corresponding Floquet state
 $|\phi_{\alpha,k}\>$ with an initial wave packet $|\psi\>$. In (a) this
 wave packet is a coherent state located in the chaotic part of the phase space
 of a single unit cell, in (b) it is concentrated on a torus inside the major
 regular island (cf Fig.~\protect\ref{psecmap}a).}
\end{figure}
Consider now a wave packet localized initially within a single unit cell and,
inside this unit cell, on one of the invariant sets of the classical dynamics.
Then the weights $|\psi_{\alpha,k}|^{2}$ are approximately homogeneous in $k$
but concentrated on the diabatic bands corresponding to the supporting
invariant set. This is illustrated in Fig.~\ref{fig:bw}.  Consequently, the
asymptotic velocity is an average over the corresponding band slopes. For
example, for a wave packet started inside the chaotic sea we expect a
value close to the classical chaotic transport velocity because this
is the average slope of the chaotic bands, see Eq.~(\ref{qvch}).  We
confirm this semiclassical result in Fig.~\ref{fig:wpvel}, where the
average position of two chaotic wave packets is shown over a large time
interval and for two different values of $N=h^{-1}$. In agreement with
Eq.~(\ref{wptrans}), we observe a linear dependence on time with very small
fluctuations, i.e., asymptotically there is indeed directed ballistic quantum
transport. The precise value of the velocity depends on the initial conditions
but these fluctuations decrease with $h$ and the average approaches the
classical transport velocity. Typically the quantum velocity for a
semiclassical chaotic wave packet is slightly above the classical value. This
is a consequence of the hierarchical phase-space regions around the embedded
islands which communicate with the main chaotic sea only via leaky cantori.
Depending on $h$, quantum transitions across some of these cantori are
possible only by tunneling, i.e., they are almost blocked. Therefore the part
of the chaotic component enclosed by these cantori effectively
belongs to the regular island \cite{K+00} and, according to the sum rule
(\ref{SumRule2}), this enhanced island size is compensated by a correspondingly
larger chaotic transport velocity.
\begin{figure}[htb]
 \def\fw{6cm}
 \centerline{\psfig{figure=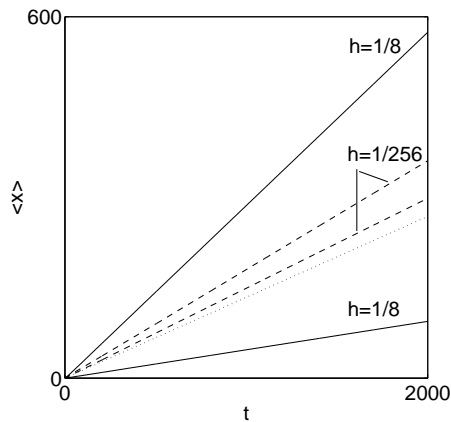,width=\fw}}
 \caption{\label{fig:wpvel} Position vs time for wave packets initialized as
 coherent states inside the chaotic part of phase space of the minimal ratchet
 model Eq.~(\protect\ref{mfomap}). Two different values of $h$ and two
 different initial conditions are used.  The dotted line shows the classical
 chaotic transport velocity.  }
\end{figure}

\subsubsection{Dynamical tunneling}
\label{sec:dyntun}
On first sight it may surprise that the division of classical phase space into
invariant sets can influence the long-time quantum dynamics. After all,
classically impenetrable barriers can be crossed in quantum dynamics by
tunneling. Tunneling is known best for the case of energetic barriers, e.g.,
in a double-well potential. Dynamical tunneling is the generalization of this
phenomenon to barriers in phase space \cite{DH81} and was recently
demonstrated experimentally \cite{SOR01,H+01c}. If in quantum dynamics no
strict barriers exist, the wave packet should explore the entire accessible
phase space for sufficiently long time and consequently directed transport
should vanish, at least on average. We have seen in the previous section that
this is not the case. So what is the role of tunneling in Hamiltonian ratchets?

To answer this question we consider a wave packet which is initially prepared
inside the regular island within the unit cell $\nx=0$. Classically, such an
initial distribution is simply transported along the chain of regular islands with
a velocity corresponding to the winding number $w_{r}$, i.e., $P_{r}(x+w_{r}
t,t)=P_{r}(x,t=0)$. This property is conserved in the quantum dynamics if we
neglect the narrow avoided crossings in the band spectrum which account for
the difference between adiabatic and diabatic bands. Let us demonstrate this
for the regular island in our minimal model which has winding number
$w_{r}=-1$. The diabatic regular bands are straight lines with slope $w_{r}$,
i.e.,
\begin{equation}
\epsilon_{r,k}=\epsilon_{r,0}+w_{r}k\,.
\end{equation}
As illustrated in Fig.~\ref{fig:bw}b, a localized initial wave packet can be
constructed from such a band by a uniform superposition of all states
\begin{equation}
\Psi(x,t=0)=\int_{0}^{1}{\rm d}k\,\phi_{r,k}(x)\,.
\end{equation}
We restrict attention to times which are a multiple of the period $\wt_r$ of
the central orbit inside the island. Then $w_{r}\,t$ is an integer which
indicates one particular unit cell. We measure $x$ relative to that unit cell 
and find for the wave packet
\begin{eqnarray}
\Psi(x+w_{r}\,t,t)&=&
\int_{0}^{1}{\rm d}k\,\exp(-2\pi\i\epsilon_{{\rm r},k}t)\,
\phi_{r,k}(x+w_{r}\,t)
\nonumber\\&=&
\int_{0}^{1}{\rm d}k\,\exp(2\pi\i(kw_{r}-
\epsilon_{r,k})t)\,\phi_{r,k}(x)
\nonumber\\&=&\exp({-2\pi\i\,\epsilon_{r,0}t})\,
\Psi(x,0)\,.
\end{eqnarray}
This shows that the wave packet is indeed transported like the corresponding
classical distribution. It has the asymptotic velocity $w_{r}$ and does
not show any spreading, i.e., there is no signature of dynamical tunneling
within the approximation of diabatic bands.

\begin{figure}[htb]
 \centerline{
  \psfig{figure=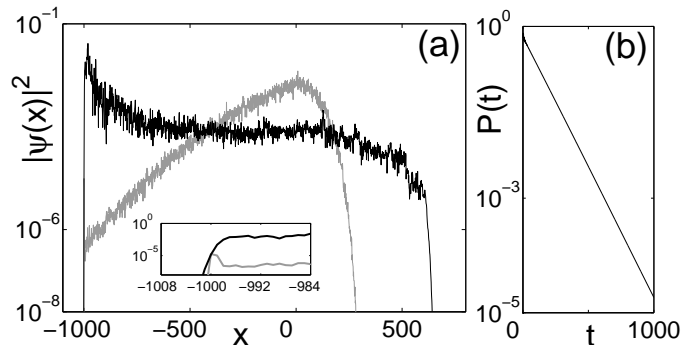,width=90mm}
 }
 \caption{\label{fig:longtimewp}
   (a) Black line: Wave packet prepared in the regular island of the unit cell
   $x=0$ and propagated to time $t=1000$ in the minimal ratchet model
   Eq.~(\protect\ref{mfomap}) with $h^{-1}=16$. The classical probability
   would be restricted to the unit cell $x=-1000$, while the quantum wave
   packet has tunneled out of this ``classical'' unit cell and starts
   spreading. However, there is a large peak lacking slightly behind
   the classically
   expected position. Gray line: Same for irrational $h^{-1}=16+\sigma$. In
   this case the Floquet operator has no spatial periodicity. The part of the
   wave packet outside the classical unit cell localizes and develops an
   asymmetric envelope with approximately exponential tails.  Inset: The
   probability to remain inside the classically expected unit cell $x=-1000$
   is the same for rational and irrational $h$.  (b) Due to dynamical
   tunneling the quantum probability in this ``classical" unit cell decays
   exponentially as a function of time. With respect to this decay the
   periodic model with $h^{-1}=16$ is almost indistinguishable from the
   aperiodic model with irrational $h^{-1}=16+\sigma$.
 }
\end{figure}

We conclude hat tunneling out of an island in classical phase space is encoded
in the avoided crossings between the regular and the chaotic bands.  These
avoided crossings show up in the regular bands as deviations from the straight
line $\epsilon_{r,0}+\omega_{\rm r}k$. Close to an avoided crossing the regular
bands are bent towards the chaotic bands, i.e., the actual slope is
$k$-dependent and slightly smaller than $w_{r}$. Using this qualitative
information about the shape of the regular bands we can make a prediction for
the shape of the wave packet at very large times $t\to\infty$. In this regime
the wave packet can be calculated from Eq.~(\ref{tdwp}) in stationary phase
approximation. We find
\begin{eqnarray}\label{longtimewp}
\Psi(X+x,t)&=&\int_{0}^{1}{\rm d}k\,
\exp(2\pi\i(kX-\epsilon_{r,k}t))\,\phi_{r,k}(x)\nonumber\\ &=&
\sum_{\epsilon'_{r,k}={X/t}}\sqrt{\i/|\epsilon''_{r,k}t|}\\
&&\times\exp(2\pi\,\i\,[k\epsilon'_{r,k}-\epsilon_{r,k}]t)\,
\phi_{r,k}(x)\,.
\nonumber
\end{eqnarray}
We have again decomposed position into a large integer $X$ denoting the unit
cell and the remaining fraction $0<x<1$. $\phi_{r,k}(x)$ is considered a
slowly varying pre\-factor of the rapidly oscillating phase. The points of
stationary phase in Eq.~(\ref{longtimewp}) select the Bloch states whose
superposition yields the wave packet at time $t$ and position $X$. It is no
surprise that these are exactly the points for which the slope of the band
corresponds to the velocity $X/t$. Due to avoided crossings, the actual slope
of the regular bands is smaller than $w_{r}$. Hence for the transition to
the unit cell $X=w_{r}t$ where all classical probability is concentrated, no
points of stationary phase with real $k$ exist: To leading order this
process is forbidden in quantum mechanics! There might be complex solutions of
the equation $\epsilon'_{r,k}=w_{r}$, but then the exponent in
Eq.~(\ref{longtimewp}) has a real part and the contribution will be
exponentially small in $t$, which is indeed observed in
Fig.~\ref{fig:longtimewp}(b). The main part of the wave packet is concentrated
not in the ``classical'' unit cell but rather at positions for which real
points of stationary phase exist in Eq.~(\ref{longtimewp}).  These correspond
to velocities distributed narrowly around a value 
slightly below the classical velocity. Due to this dispersion in the
velocities, induced by avoided crossings, the wave packet will spread
ballistically in time and will be peaked behind the classically expected
position (Fig.~\ref{fig:longtimewp}(a)).

For a wave packet initially prepared in the chaotic part of a unit cell the
influence of tunneling is much less pronounced (not shown): Although the
narrow avoided crossings with regular bands do modify the chaotic bands as
well, the existence of points of stationary phase in an expansion similar to
Eq.~(\ref{longtimewp}) is unaffected: due to the wide avoided crossings
between themselves, the chaotic states have a large variation in their
velocities around the classical value anyway.

We have thus identified the r\^ole of tunneling in Hamiltonian ratchets.  It
leads to avoided crossings between regular and chaotic states (or between
regular states with different winding numbers). In the dynamics of initially
localized wave packets tunneling shows up mainly in the evolution of regular
states, which slightly lag behind the position expected from classical
considerations. We stress again that tunneling is not able to hinder directed
ballistic transport of such wave packets even for infinite time.

An interesting and important special case are systems with a symmetry-related
pair of counter-moving regular islands like the kicked rotor in presence of
accelerator modes. Dynamical tunneling between such island pairs was
demonstrated experimentally \cite{SOR01,H+01c}. It is crucial to understand
the difference between our argumentation above and this situation.
First we note that a pair of symmetry-related islands is {\em not} analogous
to a symmetric double-well potential. In the latter case all eigenstates are
superpositions of left and right. Below the barrier top, their
eigenenergies form quasidegenerate doublets and thus contribute
to tunneling. In the case of counter-moving islands this applies only
to the vicinity of avoided crossings between the corresponding bands
where indeed they form a doublet. Away from these isolated and
semiclassically small regions in $k$-space the bands are approximately
straight lines but {\em with opposite slopes}, i.e., there is no
systematic degeneracy. In this paper we consider wave
packets initially localized inside one unit cell. In $k$-space such a
wave packet is extended. Therefore its weight in the vicinity of
avoided crossings, where it contributes to tunneling, is negligible. By
contrast, in the experiments mentioned above the wave packets
extend initially over many unit cells. Therefore, in $k$-space they
may well be concentrated right at avoided crossings. Then, and only
then, dynamical tunneling is the expected consequence.


\subsection{Quantum transport in the presence of disorder}
\label{sec:disorder}

\begin{figure}[htb]
 \centerline{
  \psfig{figure=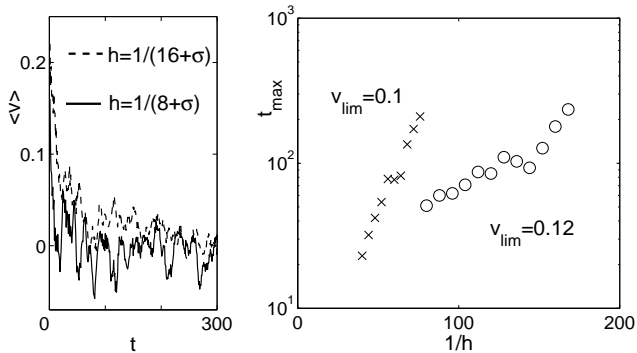,width=8.5cm}
 }
 \caption[tmax]{(left) Averaged velocity expectation values of wave
        packets initialized in the classically chaotic region at
        $(x_0 , p_0 )=(0,0.25\pm 0.05)$. Beyond $t_{\rm loc}$, the velocity
        oscillates around zero.
        (right) Time $t_{\rm max}$ at which the averaged velocity
        expectation value falls short of a given limit under variation of $h$.
        For numerical reason, we chose two different values for $v_{\rm lim}$.
        }
\label{fig:tmax}
\end{figure}

In this last section we will describe some modifications of the quantum
transport in a situation, when the exact quantum periodicity is destroyed by
weak static disorder. As explained above in Section~\ref{sec:floqop} this can
be realized easily within our minimal ratchet model by choosing an irrational
value of $h$. In this case the Bloch theorem does not apply anymore and, on a
large scale, we expect dynamical localization of wave packets and
eigenstates. The properties of the eigenstates and in particular the failure
of the semiclassical eigenfunction hypothesis in this case have been studied
in \cite{H+02}. We will here concentrate on the evolution of wave packets in
the presence of disorder. In Fig.~\ref{fig:longtimewp} (gray line) we display
the shape of a wave packet which was initialized in the regular island of unit
cell $X=0$ at time $t=1000$. Initially the wave packet follows the classical
evolution, i.e., it is transported at velocity $v=-1$ and looses probability
due to tunneling. The process of tunneling out of the island is essentially
the same as in the case of a periodic system with rational $h$. This is
demonstrated by Fig.~\ref{fig:longtimewp}b and also by the inset of
Fig.~\ref{fig:longtimewp}a, where one can see that the probability remaining
inside the classical unit cell is the same for both systems. However, the fate
of the probability which has tunneled out of the island is entirely different
from the periodic case. We see in Fig.~\ref{fig:longtimewp}a that the wave packet
develops exponential tails which are characteristic of localization. Unlike
the periodic case, the maximum of the wave packet is not close to the classical
expectation but rather close to the origin, i.e., the disorder prevents
quantum transport despite the underlying classical ratchet mechanism. The
latter is manifest, however, in the asymmetric shape of the wave
packet which has a much longer tail in the direction of classical transport.

Similarly, disorder does also affect wave packets which are initialized in the
chaotic sea. Fig.~\ref{fig:tmax}a shows the velocity expectation value for
such a wave packet at two different values of the effective Planck's constant
$h$. There is an initial period when $\<v\>\sim v_{\rm ch}$, but then the
velocity drops to zero because the wave packet tunnels into the island and
finally occupies the whole available phase space. The time for this process is
expected to scale as $t\sim\e^{c/h}$ \cite{HOA84}. As Fig.~\ref{fig:tmax}b
shows, this is also the time scale for which the quantum ratchet shows
transport in the presence of disorder.  This maximum ratchet operation time
$t_{\rm max}$ can be defined as the time at which the velocity of a wave
packet falls below a certain threshold. In Fig.~\ref{fig:tmax}b $\log t_{\rm max}$ is seen to
be approximately proportional to $h^{-1}$. Hence, in the deep semiclassical regime
the quantum ratchet can work over an exponentially long time even in the
presence of static disorder.

\section{Discussion}
\label{sec:disc}
The study of ratchets has largely been motivated by the interest in
the physical principles of intracellular transport: Motor molecules,
driven by chemical energy, are moving along chain molecules whose length
is of the order of the cell size, and which consist of millions of
units concatenated in a highly ordered manner, resembling the crystal order
encountered in inorganic solids. It is therefore natural to
model them as one-dimensional, infinitely extended potentials with exact
spatial translation invariance, but with reflection symmetry manifestly
broken to define a preferred direction of transport. 

While the breaking of mirror symmetry is crucial to obtain directed transport,
the r\^ole of translation invariance appears circumstantial, at most of
heuristic importance for the theoretical description. Translation invariance
has been indispensable, however, in order to achieve first analytical and
numerical results on directed transport in ratchets. In the present context of
Hamiltonian systems, it allowed us to show that directed transport comes about
by counter propagating phase-space flows within regular and chaotic components
of systems with a mixed phase space. Moreover, quantum ratchets
are obtained by quantizing Hamiltonian ratchets in the framework of Bloch
theory; they exhibit transport at similar rates as their classical
counterparts, at least in the semiclassical regime.

Real systems showing directed transport, biological or physical, though,
break translation invariance in various ways and to various degrees, the
only exception being systems where the spatial coordinate is cyclic,
like in biological ``rotation motors'' or in pumping devices in a closed
configuration \cite{Coh03}. In the following we discuss a number of
typical deviations from spatial periodicity and their consequences for
transport. Since quantum systems are far more sensitive to the presence or
absence of symmetries than classical ones, the question concerning
imperfections of translation invariance becomes even more crucial
on the quantum level. 

Experimental realizations of Hamiltonian ratchets, as in optical lattices or
in solid-state devices, always show a certain amount of \emph{disorder}, in the
form of small stochastic differences between the unit cells. Classically,
smooth spatial disorder, if it is not too strong, will not completely disrupt
the phase-space structures underlying transport on short time scales, thus
it has only a minor effect on transport \cite{APD03}. For long times, however,
we expect that transport is destroyed. In extended quantum systems arbitrarily
weak randomness in the potential immediately leads to localization. As we show
in Section \ref{sec:disorder}, even a type of disorder that is invisible in
the classical dynamics entails a breakdown of quantum transport on a timescale
proportional to the localization length. It should be kept in mind, however,
that localization as a quantum coherence effect is counteracted, in turn, by
incoherent processes caused by the unavoidable coupling to ambient degrees of
freedom, or similarly by a ``noisy'' driving that breaks \emph{temporal}
periodicity. While it is well known that in this way, incoherence partially
restores diffusive transport in systems with dynamical localization
\cite{DG90}, its effects on \emph{directed} transport remain to be explored.

The presence of a \emph{spatially homogeneous force} breaks translational
invariance in a more controlled yet radical manner. Rather than forming an
unavoidable nuisance, it may be imposed intentionally to extract work from a
ratchet. Moreover, it allows to define a \emph{stall force} as the external
bias just sufficient to bring transport to a standstill \cite{RH02},
and to ascribe an efficiency to ratchets. In contrast to disorder, a
finite mean potential gradient forms a perturbation of unbounded
amplitude, and thus radically changes the structure of the classical
ratchet phase space. Still, as explained in Section \ref{sec:bias},
directed regular transport reacts smoothly on an external bias, i.e.,
it requires a gradient of the order of those present in the original
periodic potential to be completely suppressed. On the quantum level,
additional complications arise in that eigenstates become metastable
and eigenenergies correspondingly complex.  This situation can be
handled in a framework similar to scattering theory \cite{GKK02}. Its
application to ratchets is under way. 

Finally, in most physical setups, transport takes place between two
``terminals'', typically modelled as electron reservoirs. This amounts to
confining the ratchet proper to a finite section of space---yet another
elementary way to break translational symmetry. Taking it into account
would allow to make contact with a different, but closely related paradigm
of directed transport: \emph{Pumps} are devices that channel a
well-defined amount of charge, mass, etc., per cycle of an applied force
from one terminal to the other \cite{Bro98,A+00}. Obviously, pumps can be
considered as ratchets reduced to a finite number of unit cells, or
conversely, ratchets could be constructed by concatenating an infinite
number of pumps or equivalently, by closing the pumping circuit. The
only difference lies in the kind of model usually studied in these
respective contexts, namely fast drivings resulting in a chaotic
dynamics in one case, slowly driven potential wells that resemble
peristaltic pumps in the other \cite{Bro98}. But this is an artificial
distinction: It has been shown recently that driven chaotic scattering
systems, employed as pumps, also generate directed transport if all
relevant binary symmetries are broken \cite{DGS03}. 

In order to study ratchets as realistic devices clamped between reservoirs
at given temperatures and chemical potentials, however, another crucial
building block is missing, a quantum statistical theory of transport under
strong time-dependent driving far from equilibrium. For first approaches
to this problem from the points of view of quantum scattering and quantum
transport theory, see Refs.~\cite{MB02} and \cite{Coh03}, respectively.
\\[5mm]

\acknowledgments We thank Marc-Felix Otto for his contribution to the project
and in particular for the numerical data underlying
Figs.~\ref{fig:velo}--\ref{fig:tmax}. We benefitted from discussions with D.~Cohen,
S.~Fishman, S.~Flach, P.~H{\"a}nggi, M.~Holthaus, P.~Reimann, O.~Yevtushenko.
TD thanks for the hospitality enjoyed during stays at MPI f\"ur Physik
komplexer Systeme (Dresden), MPI f\"ur Str\"omungsforschung (G\"ottingen),
Universit\"at Kaiserslautern, Centro Internacional de Ciencias (Cuernavaca),
Weizmann Institute of Science (Rehovot), and Technion -- Israel Institute of
Technology (Haifa). Moreover, TD acknowledges financial support by the
Volkswagen Foundation (project I/78235), by the Weizmann Institute through a
Weston Visiting Professorship, by Colciencias (project 310223), and by
Direcci\'on Nacional de Investigaci\'on (DINAIN, project DI00C1255) and
Divisi\'on de Investigaci\'on, Sede Bogot\'a (DIB, project 803684) of
Universidad Nacional de Colombia.

\appendix

\section{Change of mean momentum of a KAM torus}\label{notorus}
In this Appendix we consider a non-contractible KAM torus that can be
specified by the functional dependence of the momentum on position and
time $p(\x,t)$. Note that the existence of such a function is an
assumption which simplifies our reasoning.

We consider the average of the function $p(\x,t)$ along the torus and replace
the integral representation of this quantity, Eq.~(\ref{meanp}), by a
Riemann sum over $N\to\infty$ discrete points $\x_{n}=n/N$,
$p_{n}=p(\x_{n},\t)$
\begin{eqnarray}\label{discrete}
\overline p(\t)\approx \sum_{n=1}^{N} (\x_{n+1}-\x_{n}) p_{n}\,.
\end{eqnarray}
Similarly we introduce a discrete time increment $\delta\t$ and find that
the $N$ phase space points $(\x_{n},p_{n},\t)$ in Eq.~(\ref{discrete}) evolve
to $(\tilde\x_{n},\tilde p_{n},\t+\delta\t)$ with
\begin{equation}
\tilde\x_{n}=\x_{n}+p_{n}\delta\t\qquad
\tilde p_{n}=p_{n}-V'(\x_{n},\tau)\delta\t\,.
\end{equation}
Now we use these new points to discretize the integral representing $\overline
p(\t+\delta\t)$. In this way we obtain an expression for the time derivative
of $\overline p$, which we evaluate to leading order in $\delta\t$ and
$N^{-1}$ and then transform back to an integral. We obtain
\begin{eqnarray}
{\d\over \d\t}\overline p(\t)&\approx& {1\/\delta\t}\sum_{n=1}^{N} 
[(\tilde\x_{n+1}-\tilde\x_{n}) \tilde p_{n}-(\x_{n+1}-\x_{n}) p_{n}]
\nonumber\\&=&{1\/\delta\t}\sum_{n=1}^{N} 
\Big\{\Big[{1\/N}+(p_{n+1}-p_{n})\delta\t\Big]
\nonumber\\&&\qquad\quad
\times\Big[p_{n}-V'(\x_{n},\tau)\delta\t
\Big]-{1\/N}\,p_{n}\Big\}
\nonumber\\&\approx&
\sum_{n=1}^{N} 
\Big\{-{1\/N}V'(\x_{n},\tau)+(p_{n+1}-p_{n})\,p_{n}\Big\}
\nonumber\\&\approx&
{1\/N}\sum_{n=1}^{N} 
\Big\{-V'(\x_{n},\tau)+p_n\,p'(\x_{n},\t)\Big\}
\nonumber\\&\approx&
\int_{0}^{1}\d\x\Big\{-V'(\x,\tau)+p(\x,\t)\,{\partial\over \partial\x}p(\x,\t)\Big\}
\nonumber\\&=&
\int_{0}^{1}\d\x
\Big\{-V'(\x,\tau)+{1\over 2}{\partial\over \partial\x}p^{2}(\x,\t)\Big\}
\nonumber\\&=&
-\int_{0}^{1}\d\x\,V'(\x,\tau)
\end{eqnarray} 
For the last line we have used the periodicity of the function $p(\x,\t)$ with
respect to $\x$. By integration with respect to $\tau$ we find
Eq.~(\ref{incr}) which was the purpose of this Appendix.


\section{Generalized form factor for an island chain}\label{app:ffreg}
\def\wt{\mu} \def\wx{\nu} In this appendix we derive Eq.~(\ref{kr}).  We
consider the contribution to the form factor from one particular chain of
regular islands $r$. If the winding number is $w^{\rm cl}_{r}=\wx_{r}/\wt_{r}$
then inside a unit cell this island chain consists of $\wt_{r}$ islands which
are traversed in sequence. In the semiclassical limit, we associate (diabatic)
bands with index $\alpha$ to the island chain. These bands consist of straight
line segments with the slope $w^{\rm q}_{r}=w^{\rm cl}_{r}$, cf.\ 
Eq.~(\ref{qwinding}). The segments are connected such that the diabatic band
as a whole is periodic in $\epsilon$ and $k$, with periods $\wx_{r}$
and $\wt_{r}$, respectively. It is easy to see that for a
given value of $k$ there are $\wt_{r}$ equidistant segments (values of the
quasienergy) pertaining to the same diabatic band $\alpha$.  Semiclassically,
the number of states associated to the island chain for given $k$ is
approximately $f_{r} N$ where $f_{r}$ is the fraction of phase space occupied
by the island chain as a whole and $N=h^{-1}$ is the total number of bands. It
follows that the number of complete diabatic bands associated with the island
is ${f_{r}N/\wt_{r}}$.

To integrate over a diabatic band consisting of many straight segments it is
convenient to consider instead an extended Brillouin zone in which the band
corresponds to a single straight line
\begin{equation}\label{regband}
\epsilon_{r,\alpha,k}=
\left(\epsilon_{r,\alpha,0}+\frac{\wx_r}{\wt_r}k\right)\mod 1\,,\qquad k\in
[0,\wt_r)\,.
\end{equation}
In this way we can perform the $k$-integration in Eq.~(\ref{spamp}) and find
\begin{eqnarray}\label{kintreg}
u_{\alpha}({\nx},{\mt})&=&\int_{0}^{\wt_{r}}{\rm d}k\,
\e^{\left[2\pi\i\left(k{\nx}-\left(\epsilon_{r,\alpha,0}+
\frac{\wx_r}{\wt_r}k\right){\mt}\right)\right]}
\nonumber\\&=&\wt_r\,\e^{-2\pi\i\epsilon_{r,\alpha,0}{\mt}}
\int_{0}^{1}\d\kappa\,\e^{2\pi\i(\wt_r{\nx}-\wx_r{\mt})\kappa}
\nonumber\\&=&\wt_r\,\e^{-2\pi\i\epsilon_{r,\alpha,0}{\mt}}
\delta_{\wt_r{\nx}-\wx_r{\mt}}\,.
\end{eqnarray} 
For the contribution of the island chain $r$ to the form factor we have
now 
\begin{eqnarray}\label{ffregsum}
&&K_{r}({\nx},{\mt})={1\over N}
\\\nonumber&&\qquad\times
\Bigg\langle\Bigg|
\sum_{\alpha=1}^{N\,f_{r}/\wt_r}\wt_r
\e^{-2\pi\i\epsilon_{r,\alpha,0}{\mt}}
\delta_{\wt_r{\nx}-\wx_r{\mt}}\Bigg|^2\Bigg\rangle\,,
\end{eqnarray}
i.e., we have to perform a sum over quasienergies at fixed Bloch number $k=0$
which can be done in the same way as for the spectrum of eigenenergies
pertaining to regular states of an autonomous system \cite{BT77,D96}. We
assume the dynamics within the island to deviate sufficiently from harmonic
vibrations around its central orbit. Then the spectrum of quasienergies
$\epsilon_{r,\alpha,0}$ will not be equidistant and the phases in
Eq.~(\ref{ffregsum}) from different $\alpha$ can be assumed uncorrelated in the
semiclassical limit. This allows to replace $|\sum_{\alpha}\ldots|^{2}$ by the
number of terms in the sum, which finally yields Eq.~(\ref{kr}).

\section{Wave Packet transport}\label{app:wt}
We compute the average position of a wave packet $\psi(x,t)$
for long time $t\gg 1$. First we write the wave packet as a superposition of
Floquet eigenstates $\phi_{\alpha,k}(x)$ with quasienergy
$\epsilon_{\alpha,k}$
\begin{eqnarray}
\psi(x,t)&=&
\sum_{\alpha}\int_{0}^{1}{\rm d}k\,
 \psi_{\alpha,k}(t)\,\phi_{\alpha,k}(x)
\nonumber\\&=&
\sum_{\alpha}\int_{0}^{1}{\rm d}k\,
 \psi_{\alpha,k}\,\e^{-2\pi\i\epsilon_{\alpha,k}t}\,\phi_{\alpha,k}(x)\,,
\label{psixt}
\end{eqnarray}
where
\begin{equation}
\psi_{\alpha,k}=\int_{-\infty}^{+\infty}
{\rm d}x\,\phi^{*}_{\alpha,k}(x)\,\psi(x,t=0)\,.
\end{equation}
The integral representing the expectation value of $\hat x$ for the wave packet
(\ref{psixt}) can be split into two contributions $\x$, $X$
corresponding to length scales within a unit cell and over many unit cells,
respectively
\begin{eqnarray}
x(t)&=&\int_{-\infty}^{+\infty}{\rm d}x\,x|\psi(x,t)|^2
\nonumber\\&=&
\int_{0}^{1}{\rm
d}x\,\sum_{n=-\infty}^{+\infty}(x+n)\,\left|\psi(x+n,t)\right|^2
\nonumber\\&=&
\x(t)+X(t)\,.
\end{eqnarray}
Naturally, the contribution from the dynamics inside the unit cells is bounded
from above by the size of the unit cell
\begin{eqnarray}
\x(t)&=&
\int_{0}^{1}{\rm d}x\,x\sum_{n=-\infty}^{+\infty}\,\left|\psi(x+n,t)\right|^2
\nonumber\\&\le&
\int_{0}^{1}{\rm d}x\sum_{n=-\infty}^{+\infty}\,\left|\psi(x+n,t)\right|^2
\nonumber\\&=& 1
\end{eqnarray}
(the last equality expresses the normalization of the wave packet).
Therefore $\x$ is irrelevant for directed ballistic transport.

Evaluating the term that describes the wave packet on large scales, we use the
Bloch theorem to switch from position representation to the conjugate variable
$k$, where a spatial shift corresponds to differentiation. We have
\begin{eqnarray}
n\psi(x+n,t)&=&n\,\sum_{\alpha}\int_{0}^{1}{\rm
d}k\,\psi_{\alpha,k}(t)\,\phi_{\alpha,k}(x+n)
\nonumber\\&=&
\sum_{\alpha}\int_{0}^{1}{\rm d}k\,\psi_{\alpha,k}(t)\,\phi_{\alpha,k}(x)\,
n\e^{2\pi\i\,kn}
\nonumber\\&=&
\sum_{\alpha}\int_{0}^{1}{\rm d}k\,\psi_{\alpha,k}(t)\,\phi_{\alpha,k}(x)\,
{{\rm d}\/{\rm d}k}\,{\e^{2\pi\i kn}\/2\pi\i}
\nonumber\\&=&
-\sum_{\alpha}\int_{0}^{1}{\rm d}k\,{\e^{2\pi\i kn}\/2\pi\i}
{{\rm d}\/{\rm d}k}\psi_{\alpha,k}(t)\,\phi_{\alpha,k}(x)\,.
\nonumber
\end{eqnarray}
The last line follows from partial integration and the periodicity in $k$
of $\e^{2\pi\i kn}\psi_{\alpha,k}(t)\,\phi_{\alpha,k}(x)$. Inserting this
into
\begin{eqnarray}
X(t)&=&
\int_{0}^{1}{\rm d}x\sum_{n=-\infty}^{+\infty}\,n\,\left|\psi(x+n,t)\right|^2
\nonumber
\end{eqnarray}
and decomposing also the complex conjugate $\psi^{*}(x+n,t)$ into Floquet
states we find
\begin{eqnarray}
X(t)&=&
-\int_{0}^{1}{\rm d}x\sum_{n=-\infty}^{+\infty}\,
\sum_{\alpha,\alpha'}\int_{0}^{1}{\rm d}k\,\int_{0}^{1}{\rm d}k'\,
{\e^{2\pi\i (k-k')n}\/2\pi\i}
\nonumber\\
&&\times\psi^{*}_{\alpha',k'}(t)\,\phi^{*}_{\alpha',k'}(x)
{{\rm d}\/{\rm d}k}\psi_{\alpha,k}(t)\,\phi_{\alpha,k}(x)
\nonumber\\&=&
-{1\/2\pi\i}
\int_{0}^{1}{\rm d}x\,
\sum_{\alpha,\alpha'}\int_{0}^{1}{\rm d}k\,\times
\nonumber\\
&&\psi^{*}_{\alpha',k}(t)\,\phi^{*}_{\alpha',k}(x)
{{\rm d}\/{\rm d}k}\psi_{\alpha,k}\,\e^{-2\pi\i \epsilon_{\alpha,k}t}
\,\phi_{\alpha,k}(x)\,.
\nonumber
\end{eqnarray}
The last line follows here from Poisson summation over $n$. In this expression
${\rm d}/{\rm d}k$ acts on a product of three terms, but as $t\to \infty$ the
dominant contribution comes from the derivative of the exponential. Neglecting
the other two terms which are bounded, and using the orthonormalization of
Floquet states we finally obtain Eqs.~(\ref{wptrans}),
(\ref{wpvel}).

For higher moments of the spatial distribution the argument can be
repeated and an analogous result is obtained
\begin{eqnarray}
\<(x-x(t))^{m}\>&=&t^{m}\,\sum_{\alpha}\int_{0}^{1}{\rm d}k\,
|\psi_{\alpha,k}|^{2}\({{\rm d}\epsilon_{\alpha,k}\/{\rm d}k}\)^{m}
\nonumber\\&&
+{\rm O}(t^{m-1})\,.
\end{eqnarray}
\end{document}